\documentclass[onecolumn,showpacs,aps,prd,nobibnotes,nofootinbib,floatfix]{revtex4-1}

\usepackage{bm,graphicx,dcolumn,epstopdf,epsf, latexsym,mathbbol, amssymb,amsmath,color,slashed, mathrsfs,mathcomp,simplewick}
\usepackage[center]{subfigure}
\usepackage{multirow}
\usepackage{makecell}
\usepackage[colorlinks,linkcolor=blue,citecolor=blue,urlcolor=blue]{hyperref}

\begin{document}
	\allowdisplaybreaks
	\newcommand{\bq}{\begin{equation}}
	\newcommand{\eq}{\end{equation}}
	\newcommand{\bqn}{\begin{eqnarray}}
	\newcommand{\eqn}{\end{eqnarray}}
	\newcommand{\nb}{\nonumber}
	\newcommand{\lb}{\label}
	\newcommand{\f}{\frac}
	\newcommand{\p}{\partial}
	\newcommand{\tx}{\text}
	\newcommand{\ms}{\mathscr}
	\newcommand{\lf}{\left}
	\newcommand{\rt}{\right}
	\newcommand{\PRL}{Phys. Rev. Lett.}
	\newcommand{\PLB}{Phys. Lett. B}
	\newcommand{\PRD}{Phys. Rev. D}
	\newcommand{\CQG}{Class. Quantum Grav.}
	\newcommand{\JCAP}{J. Cosmol. Astropart. Phys.}
	\newcommand{\JHEP}{J. High. Energy. Phys.}
	\newcommand{\bea}{\begin{eqnarray}}
	\newcommand{\ena}{\end{eqnarray}}
	\newcommand{\beqa}{\begin{eqnarray}}
	\newcommand{\eeqa}{\end{eqnarray}}
	\newcommand{\red}{\textcolor{red}}

	\title{Axial gravitational quasinormal modes of a self-dual black hole in loop quantum gravity}
	
	\author{Sen Yang${}^{a, b, c}$}
	\email{120220908881@lzu.edu.cn}
	
	\author{Wen-Di Guo${}^{a, b, c}$}
	\email{guowd@lzu.edu.cn, Wen-Di Guo and Sen Yang are co-first authors of this paper.}
	
	\author{Qin Tan${}^{a, b, c}$}
	\email{tanq19@lzu.edu.cn}
	
	\author{Yu-Xiao Liu${}^{a, b, c}$}
	\email{liuyx@lzu.edu.cn, corresponding author}
	
	
	\affiliation{
		${}^{a}$ Institute of Theoretical Physics $\&$ Research Center of Gravitation, Lanzhou University, Lanzhou 730000, China\\
		${}^{b}$ Key Laboratory of Quantum Theory and Applications of MoE, Lanzhou University, Lanzhou 730000, China\\
		${}^{c}$ Lanzhou Center for Theoretical Physics $\&$ Key Laboratory of Theoretical Physics of Gansu Province, Lanzhou University, Lanzhou 730000, China\\}
	\date{\today}

	\begin{abstract}
		We study the axial gravitational quasinormal modes of a self-dual black hole in loop quantum gravity.  Considering the axial perturbation of the background spacetime, we obtain the Schr\"{o}dinger-like master equation. Then we calculate the quasinormal frequencies with the Wentzel-Kramers-Brillouin approximation and the asymptotic iteration method. We also investigate the numerical evolution of an initial wave packet on the self-dual black hole spacetime. We find the quantum correction parameter $P$ positively affects the absolute values of both the real and imaginary parts of quasinormal frequencies. We derive the relation between the parameters of the circular null geodesics and quasinormal frequencies in the eikonal limit for the self-dual black hole, and numerically verify this relation.		
	\end{abstract}
	
	
\maketitle
	
	\section{Introduction}
	\renewcommand{\theequation}{1.\arabic{equation}} \setcounter{equation}{0}
	
	The first direct detection of the gravitational wave (GW) in 2015 \cite{LIGOScientific:2016aoc}  marked an all-new era of physics and astronomy \cite{Cai:2017cbj, Bian:2021ini}. The Event Horizon Telescope has taken the first picture of a supermassive object at the center of galaxy M87 \cite{EventHorizonTelescope:2019dse, EventHorizonTelescope:2019uob, EventHorizonTelescope:2019jan, EventHorizonTelescope:2019ths, EventHorizonTelescope:2019pgp,EventHorizonTelescope:2019ggy}, and the picture of the central black hole in our Milky Way \cite{EventHorizonTelescope:2022wkp, EventHorizonTelescope:2022apq, EventHorizonTelescope:2022wok, EventHorizonTelescope:2022exc, EventHorizonTelescope:2022urf, EventHorizonTelescope:2022xqj}. Human beings can observe the universe with multi-messenger, both the gravitational wave and the electromagnetic wave. Until now, the LIGO--Virgo--KAGRA collaboration has finished three observing runs and detected 90 confident GW-burst events \cite{LIGOScientific:2018mvr, LIGOScientific:2020ibl, LIGOScientific:2021usb, LIGOScientific:2021djp}. GW-bursts, emitted from the merger of binary compact objects, bring information about gravitational theories and sources and provide us with a new approach to test general relativity in the strong gravitational field \cite{LIGOScientific:2016lio, LIGOScientific:2019fpa, LIGOScientific:2020tif, LIGOScientific:2021sio}. The whole gravitational wave waveform of a GW-burst event can be divided into three parts: inspiral, merger, and ringdown. And the ringdown part can be successfully described by the black hole perturbation theory \cite{Chandrasekhar:1985kt, Maggiore:2018sht}.
	
	A black hole with perturbations is a dissipative system, and the
	eigenmodes of this system are named quasinormal modes (QNMs). The QNMs are the spectroscopy of a black hole, because the quasinormal frequencies depend only on the black hole’s parameters, while their amplitudes depend on the source exciting the oscillations \cite{Kokkotas:1999bd, Nollert:1999ji, Berti:2009kk, Konoplya:2011qq}. According to the behavior under space inversions, the gravitational perturbations of a spherically symmetric black hole can be divided into the odd (axial) parity part and the even (polar) parity part \cite{Chandrasekhar:1985kt}. As the most successful theory for gravitational interaction, general relativity has passed many astrophysical tests \cite{Weinberg:1972kfs}. In general relativity, Regge, Wheeler \cite{Regge:1957td}, and Zerilli \cite{Zerilli:1970se} first studied the odd parity and the even parity gravitational perturbations of the Schwarzschild black hole. Moncrief first studied both the odd parity and the even parity gravitational perturbations of the Reissner-Nordstrom black hole \cite{Moncrief:1974gw, Moncrief:1974ng}. And Teukolsky first studied the gravitational perturbations of the Kerr black hole \cite{Teukolsky:1972my}. To get the quasinormal frequencies for the black hole perturbation problem, numerical methods are needed to solve the eigenvalue problem. With the development of the black hole perturbation theory, more and more numerical methods were proposed, such as the Wentzel-Kramers-Brillouin (WKB) approximations \cite{Mashhoon, Schutz:1985km, Iyer:1986np, Konoplya:2003ii, Matyjasek:2017psv, Konoplya:2019hlu}, the asymptotic iteration method \cite{Cho:2011sf}, the monodromy technique \cite{Motl:2003cd}, the series solution \cite{Horowitz:1999jd}, the resonance method \cite{Berti:2009wx}, and the Leaver’s continued fraction method \cite{Leaver:1985ax}.
	
	The singularity of general relativity is a good motivation to probe new physics. It is generally believed that a complete theory of quantum gravity has no singularity. Loop quantum gravity is exactly this case \cite{Rovelli:2004tv}. In loop quantum gravity, spacetime is made up of some basic building blocks called spin networks. In the framework of loop quantum gravity, Modesto and Premont-Schwarz constructed the Reissner-Nordstrom-like self-dual black hole \cite{Modesto:2009ve, Modesto:2008im}. Many works investigated the phenomenological implications of this black hole \cite{Alesci:2011wn, Barrau:2014yka, Dasgupta:2012nk, Sahu:2015dea, Hossenfelder:2012tc, Zhu:2020tcf, Yan:2022fkr}.
	The perturbations of the self-dual black hole also have been studied in some works, which can be divided into two categories by whether using the Arnowitt-Deser-Misner (ADM) mass of the black hole as one of the parameters fixed during calculation. Fixing the parameter $M/(1+P)^2$ instead of the ADM mass of the self-dual black hole, Chen and Wang studied the QNMs of a massless scalar field \cite{Chen:2011zzi}, Santos $et ~al.$ studied QNMs of a massive scalar field nonminimally coupled to gravity \cite{Santos:2021wsw}, Cruz $et ~al.$ studied axial \cite{Cruz:2015bcj} and polar gravitational perturbations \cite{Cruz:2020emz}. But it is worth pointing out that the effective potential in Ref.~\cite{Cruz:2015bcj} cannot be reduced to the Schwarzschild black hole case when setting all loop quantum gravity parameters equal to zero. Fixing the ADM mass of the self-dual black hole, Liu $et ~al.$ studied QNMs of the massless scalar field and electromagnetic field \cite{Liu:2020ola}, and Momennia studied the QNMs of a test scalar field \cite{Momennia:2022tug}.
	
	In this work, we focus on the axial gravitational perturbation of the self-dual black hole with fixed ADM mass, because the ADM mass is the physical mass of a black hole measured in astronomical observations. Following Ref.~\cite{Modesto:2009ve}, we assume that the self-dual black hole is described by Einstein’s gravity minimally coupled to an anisotropic fluid, and derive the master equation of the axial gravitational perturbation of the self-dual black hole. This method also was used to study the gravitational perturbations of nonsingular black holes in conformal gravity \cite{Chen:2019iuo} and non-singular Schwarzschild black holes in loop quantum gravity \cite{Bouhmadi-Lopez:2020oia}. Then, we calculate the corresponding quasinormal frequencies with the WKB approximation and the asymptotic iteration method. The influence of the quantum correction parameter $P$ on the QNMs is also studied. We find that the parameter $P$ has a positive effect on the absolute values of both the real part and the imaginary part of quasinormal frequencies, which is consistent with the conclusions for the QNMs of the scalar field and the electromagnetic field on the self-dual black hole with fixed ADM mass during calculating \cite{Liu:2020ola, Momennia:2022tug}.   Assuming the perturbation is a Gaussian packet, we investigate the numerical evolution of an initial wave packet on the self-dual black hole. Besides, Cardoso, Lemos, and Yoshida found that, in the eikonal limit, quasinormal modes of a stationary, spherically symmetric, and asymptotically flat black hole in any dimension are determined by the parameters of the circular null geodesics \cite{Cardoso:2008bp}.~We obtain the relation between the quasinormal frequencies in the eikonal limit of the axial gravitational perturbation and the parameters of the circular null geodesics in the self-dual black hole, and numerically verify this relation. The numerical results show that the relation between the parameters of the circular null geodesics and quasinormal frequencies in the eikonal limit is right in the self-dual black hole in loop quantum gravity.
	
	This paper is organized as follows. In Sec. \ref{sec2}, we derive the master equation of the axial gravitational perturbation of the self-dual black hole. In Sec.~\ref{sec3}, we calculate the corresponding quasinormal frequencies with the WKB approximation method and the asymptotic iteration method. And we investigate the numerical evolution of an initial wave packet on the self-dual black hole spacetime. Then we obtain the relation between the parameters of the circular null geodesics and quasinormal frequencies in the eikonal limit in the self-dual black hole, and numerically verify this relation in Sec. \ref{sec4}. Finally, the conclusions and discussions of this work are given in Sec. \ref{sec5}.
	
	\section{Gravitational perturbation of loop quantum black hole} \label{sec2}
	\renewcommand{\theequation}{2.\arabic{equation}} \setcounter{equation}{0}
	
	The line element of the spherically symmetric self-dual black hole in loop quantum gravity is \cite{Modesto:2009ve}
	\bq\lb{metric}
	ds^2 = - f(r) dt^2 + \frac{dr^2}{g(r)} + h(r) \left( d \theta^2 + \sin ^2 \theta d \varphi ^2 \right) ,
	\eq
	where the functions $f(r)$, $g(r)$, and $h(r)$ have the following forms
	\bqn
	f(r) &=& \frac{(r-r_{+})(r-r_{-})}{r^4 + a^2_0}(r + r_{0})^2 , \lb{fr} \\
	g(r) &=& \frac{(r-r_{+})(r-r_{-})}{r^4 + a^2_0} \frac{r^4}{(r + r_0)^2}, \lb{gr} \\
	h(r) &=& r^2 + \frac{a^2_0}{r^2} ,
	\eqn
	where $a_0 \simeq 5 l^2_P /8\pi$ ($l_p$ is the Planck length) is related to the minimum area gap of loop quantum gravity, $r_{+} = 2M/(1 + P)^2$ is the outer (event) horizon, with $P$ a function of the polymeric parameter $
	\delta_b$ related to the geometric quantum effect of loop quantum gravity. $ r_{-} = 2 M P^2/(1 + P)^2 $ is the inner (Cauchy) horizon, $r_0 = \sqrt{r_{+} r_{-}} $, and $M$ is the ADM mass of the black hole. The deviation of the self-dual black hole from the Schwarzschild black hole is described by two quantum correction parameters $P$ and $a_0$. The constraints on the  parameter $P$ have been obtained from various astrophysical observations \cite{Sahu:2015dea, Zhu:2020tcf, Yan:2022fkr}, and the max one is $ P < 0.0675 $ \cite{Zhu:2020tcf}. Expanding Eqs.~(\ref{fr}) and (\ref{gr}) in the power of $1/r$, one can see that the maximal correction from the parameter $P$ is at the order of $(MP)/r$, while the maximal correction from $a_0$ is at the order of $a_0^2 /r^4$ \cite{Zhu:2020tcf}. In this work, we focus on the physics of QNMs outside the event horizon. And the radius of the event horizon of a typical Schwarzschild black hole with the mass of the sun is of about $3$ km, then $P \sim \mathcal{O}(10^{-2}) $ and $a_0^2 /r^4 \sim \mathcal{O}(10^{-67})$. So the effect of $a_0$ on astrophysical observation can be safely neglected, and we only care about the quantum correction from the parameter $P$.
	
\begin{figure}[h]\centering
		\includegraphics[scale =0.14]{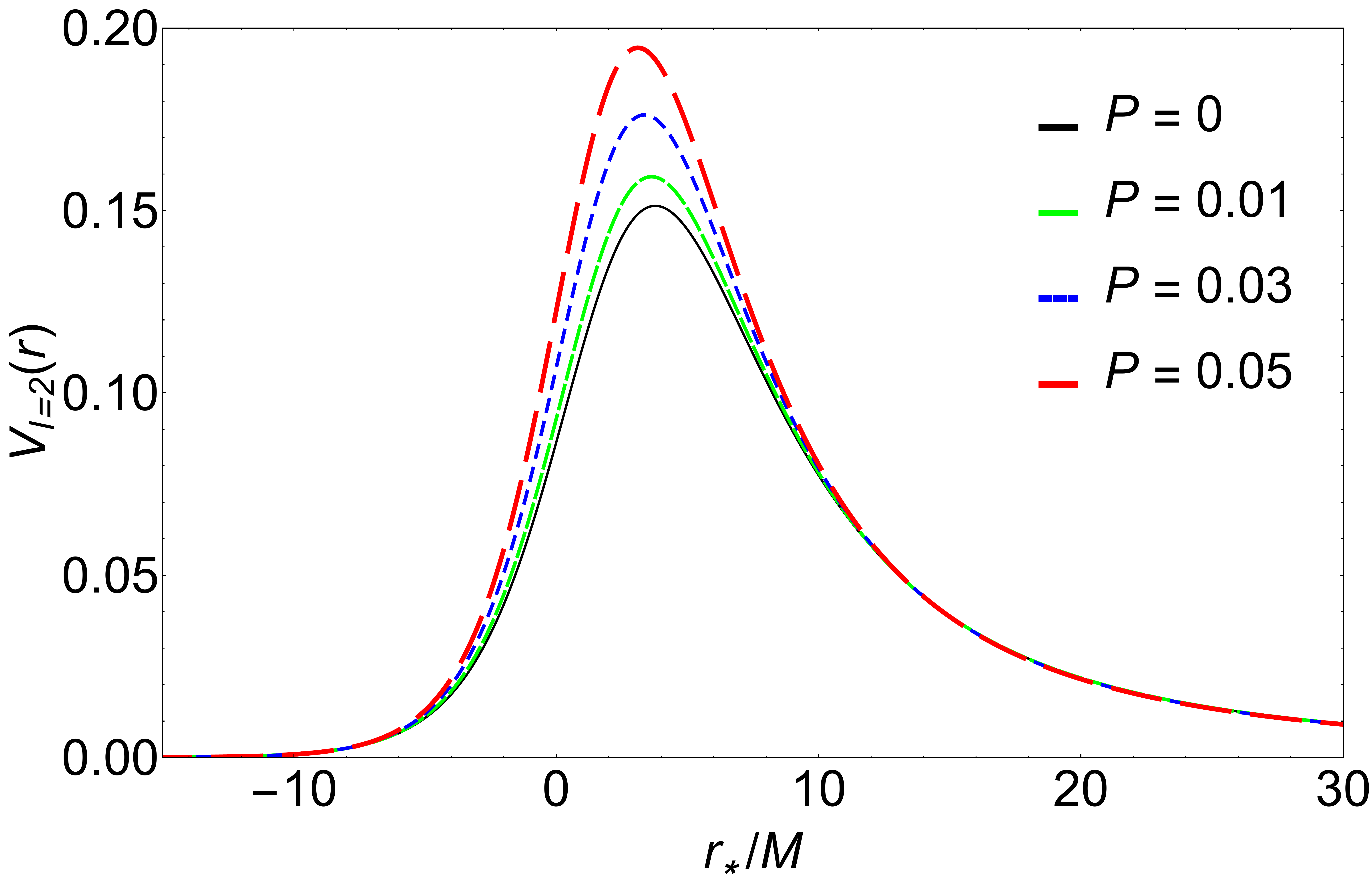}
		\caption{The effective potential (\ref{V}) in the tortoise coordinate (\ref{tortoise}) with $M = 1$, $l = 2$, and different values of the parameter $P$. The black curve shows the Regge–Wheeler potential of the Schwarzschild black hole.}
		\label{plotV}
	\end{figure}

	To study the perturbations of a spherically symmetric black hole, one can first focus on axisymmetric modes of perturbations \cite{Chandrasekhar:1985kt}. We consider a perturbed spacetime which is described by a non-stationary and axisymmetric metric as
	\bqn\lb{axisymmetric metric}
	ds^2 = -e^{2 \nu} \left( dx^0 \right)^2 + e^{2 \psi} \left( dx^1 - \sigma dx^0 - q_2 dx^2 -q_3 dx^3 \right)^2 + e^{2 \mu_2} \left( dx^2 \right)^2 + e^{2 \mu_3} \left( dx^3 \right)^2,
	\eqn
	where $\nu$, $\psi$, $\mu_2$, $\mu_3$, $\sigma$, $q_2$, and $q_3$ depend on time coordinate $t$ $(t = x^0)$, radial coordinate $r$ $(r = x^2)$, and polar angle coordinate $\theta (\theta = x^3)$. And a tetrad basis $e_{(a)}^\mu$ corresponding to the metric (\ref{axisymmetric metric}) is
	\bqn
	e^{\mu}_{(0)} &=& (e^{-\nu},~~ \sigma e^{- \nu},~~ 0,~~ 0), \nb\\
	e^{\mu}_{(1)} &=& (0,~~ e^{- \psi},~~ 0,~~ 0), \nb\\
	e^{\mu}_{(2)} &=& (0,~~ q_2 e^{- \mu_2},~~ e^{- \mu_2},~~ 0), \nb\\
	e^{\mu}_{(3)} &=& (0,~~ q_3 e^{- \mu_3},~~ 0,~~ e^{-\mu_3}).
	\eqn
	In this regard, one can project any vector or tensor field onto the tetrad frame by
	\bqn
	A_{(a)} = e^{\mu}_{(a)} A_{\mu},~~~~B_{(a)(b)} = e^{\mu}_{(a)} e^{\nu}_{(b)} B_{\mu \nu}.
	\eqn
	For a static and spherically symmetric spacetime, $\sigma$, $q_2$, and $q_3$ are zero. Then, comparing the metric (\ref{axisymmetric metric}) with (\ref{metric}), one can get
	\bqn
	e^{2 \nu} = f(r),~~~~e^{-2 \mu_2} = g(r),~~~~e^{2 \mu_3} = h(r),~~~e^{2 \psi} = h(r) \sin^2 \theta.
	\eqn
	For a self-dual black hole, one can simulate the quantum corrections with an effective anisotropic matter fluid, and write the field equation as the Einstein equation form $G_{\mu \nu} = 8 \pi T_{\mu \nu}$, where $T_{\mu \nu}$ is the effective energy-momentum tensor \cite{Modesto:2009ve}. In the tetrad frame, the field equation can be rewritten as
	\bqn
	R_{(a)(b)} - \frac{1}{2} \eta_{(a)(b)} R = 8 \pi T_{(a)(b)}.
	\eqn
	And it has been proven that the axial components of the perturbed energy-momentum tensor defined by an anisotropic fluid are zero in the tetrad formalism \cite{Chen:2019iuo}. Therefore, the master equation of the axial gravitational perturbation of the self-dual black hole can be derived from the axial components of $R_{(a)(b)}=0$. The $(1,3)$ and $(1,2)$ components of $R_{(a)(b)}|_{\text{axial}}=0 $ are
	\bqn
	\left[ h e^{ \nu - \mu_2} \left( q_{2,3} - q_{3,2} \right) \right]_{,2} &=& \left[ h e^{\mu_2 - \nu} \left( \sigma_{,3} - q_{3,0} \right) \right]_{,0}, \lb{Rab-1}\\
	\left[ h e^{ \nu - \mu_2} \left( q_{3,2} - q_{2,3} \right) \sin^3 \theta \right]_{,3} &=& \left[ h^2 e^{- \nu - \mu_2} \left( \sigma_{,2} - q_{2,0} \right) \sin^3 \theta \right]_{,0}, \lb{Rab-2}
	\eqn
	respectively, where $F_{,i} \equiv \f{\p F}{\p x^i}$. Then, one can define
	\bqn
	Q = h e^{ \nu - \mu_2} \left( q_{2,3} - q_{3,2} \right) \sin^3\theta,
	\eqn
	and rewrite Eqs. (\ref{Rab-1}) and (\ref{Rab-2}) as
	\bqn
	e^{\nu - \mu_2} \frac{Q_{,2}}{h \sin^3 \theta} &=& \left( \sigma_{,3} - q_{3,0} \right)_{,0}, \lb{RabQ-1} \\
	e^{\nu + \mu_2} \frac{Q_{,3}}{h^2 \sin^3 \theta} &=& - \left( \sigma_{,2} - q_{2,0} \right)_{,0}. \lb{RabQ-2}
	\eqn
	By differentiating Eqs. (\ref{RabQ-1}) and (\ref{RabQ-2}) and eliminating $\sigma$, one can obtain
	\bqn\lb{master eq-1}
	\frac{1}{\sin^3 \theta} \left( \frac{e^{\nu-\mu_2}}{h} Q_{,2}\right)_{,2} + \frac{e^{\nu+\mu_2}}{h^2} \left( \frac{Q_{,3}}{\sin^3\theta} \right)_{,3} = \frac{Q_{,00}}{h e^{\nu-\mu_2} \sin^3 \theta}.
	\eqn
	Considering the ansatz \cite{Chandrasekhar:1985kt}
	\bqn
	Q(r, \theta) = Q(r) Y(\theta)
	\eqn
	with $Y (\theta)$ the Gegenbauer function satisfying
	\bqn
	\frac{d}{d \theta} \left( \frac{1}{\sin^3 \theta} \frac{d Y}{d \theta}\right) = - \mu^2 \frac{Y}{\sin^3 \theta},
	\eqn
	where $\mu^2 = (l-1)(l+2)$, one can rewrite Eq. (\ref{master eq-1}) as
	\bqn
	\left( \frac{e^{\nu - \mu_2}}{h} Q_{,r} \right)_{,r} + \left( \frac{\omega^2}{h e^{\nu-\mu_2}} - \frac{e^{\nu + \mu_2} \mu^2}{h^2} \right) Q = 0.
	\eqn
	Note that here we have used the
	Fourier transformation $\p t \rightarrow - i\omega$. Then, one can define
	\bqn
	\Psi (r) = \frac{Q(r)}{\sqrt{h(r)}}.
	\eqn
	With this, we can obtain the Schr\"{o}dinger-like master equation of the axial gravitational perturbation for the self-dual black hole
	\bqn\lb{RW-eq}
	\frac{\p^2 \Psi}{\p r^2_{\ast}} + \lf[ \omega^2 -V (r) \rt] \Psi =0,
	\eqn
	where
	\bqn\lb{V}
	V(r) = \frac{f(r)(l-1)(l+2)}{h(r)} - \sqrt{f(r)g(r)h(r)}\frac{d}{dr} \left( \frac{\sqrt{f(r)g(r)}}{h(r)} \frac{d \sqrt{h(r)}}{dr} \right)
	\eqn
	is the effective potential, and $r_\ast$ is the tortoise coordinate defined by
	\bqn\lb{tortoise}
	r_{\ast} &=& \int \frac{dr}{\sqrt{f(r) g(r)}} \nb\\
	&=& r - \frac{a_{0}^2}{r_{+} r_{-}} \lf( \frac{1}{r} - \frac{r_{+} + r_{-}}{r_{+} r_{-}} \ln(r) \rt) + \frac{1}{(r_{+} - r_{-})} \lf( \frac{a^2_{0} + r^4_{+}}{r^2_{+}} \ln(r - r_{+}) - \frac{a^2_{0} + r^4_{-}}{r^2_{-}} \ln(r - r_{-})  \rt).
	\eqn
	It is worthwhile to mention that $r_{\ast}$ running from $- \infty$ to $+\infty$ matches
	$r$ from the event horizon to spatial infinity. With different values of the parameter $P$, the plots for the effective potential (\ref{V}) in the tortoise coordinate (\ref{tortoise}) are shown in Fig. \ref{plotV}. It can be seen that, the height of the effective potential increases with the parameter $P$.
	
	\section {quasinormal modes and ringdown waveforms}
	\lb{sec3}
	\renewcommand{\theequation}{3.\arabic{equation}}
	\setcounter{equation}{0}
	
	\subsection{Quasinormal Modes}
	
	The Schr\"{o}dinger-like equation (\ref{RW-eq}) has a set of complex eigenvalues, which are the QNMs of the self-dual black holes (\ref{metric}). In this work, we calculate the QNMs of the self-dual black hole using the WKB approximation method and the asymptotic iteration method.
	
	The WKB approximation method was first applied to the problem of scattering around a black hole at the first order by Schutz and Will \cite{Schutz:1985km}, and later developed to higher orders \cite{Iyer:1986np, Konoplya:2003ii, Matyjasek:2017psv, Konoplya:2019hlu}. This method can be used to solve the eigenvalue problem in which the effective potential has the form of a potential barrier and approaches to constant values at the event horizon and spatial infinity. The effective potential (\ref{V}) satisfies these conditions. And the WKB approximation works best for low overtones, i.e., modes with a long decay time, and in the eikonal limit of large $l$. Setting $a_0 =0$, $P$ from $0$ to $0.05$, and $\{l=2, 3, 4\} (0 \leq n<l)$, we use the WKB approximation to calculate the quasinormal frequencies $\omega_{nl}$ for the axial gravitational perturbation of the self-dual black hole. The results are listed in Tables \ref{Tl2}, \ref{Tl3}, and \ref{Tl4}.
	
	To use the asymptotic iteration method, we first rewrite the Schr\"{o}dinger-like equation (\ref{RW-eq}) in the $r$ coordinate as follows
	\bqn\lb{RW-r}
	f(r) g(r) \Psi''(r) + \frac{1}{2}[f'(r) g(r) + f(r) g'(r)] \Psi'(r) + [ \omega^2 - V(r) ] \Psi = 0,
	\eqn
	where the prime denotes the derivative to $r$. For the perturbation propagating in the black hole spacetime, there are two physical boundary conditions: i) $ \Psi (r_\ast) \sim e^{-i \omega r_\ast}$ as $r_\ast \rightarrow -\infty ~(r \rightarrow r_{+})$, which means the wave near the event horizon should purely enter the black hole; ii) $ \Psi (r_\ast) \sim e^{i \omega r_\ast}$ as $r_\ast \rightarrow \infty ~(r \rightarrow \infty)$, which means the wave is purely outgoing at spatial infinity. For $a_0=0$ and $P \neq 0$, the self-dual black hole has both a Cauchy horizon and an event horizon, and one can define the solution of Eq. (\ref{RW-r}) as
	\bqn\lb{solution-1}
	\Psi (r) = \frac{e^{i \omega r}}{r} (r - r_{-})^{ 1 + i \omega + i \omega r_{+}^2/(r_{+} - r_{-})} (r - r_{+})^{-i \omega r^2_{+}/(r_{+} - r_{-})} \psi(r).
	\eqn
	Taking the solution (\ref{solution-1}), we use the asymptotic iteration method to solve Eq. (\ref{RW-r}) and obtain the corresponding quasinormal frequencies with the same setting in the previous WKB calculation. The results are also listed in Tables \ref{Tl2}, \ref{Tl3}, and \ref{Tl4}.
	
\begin{figure}[h]\centering
		\includegraphics[scale =0.14]{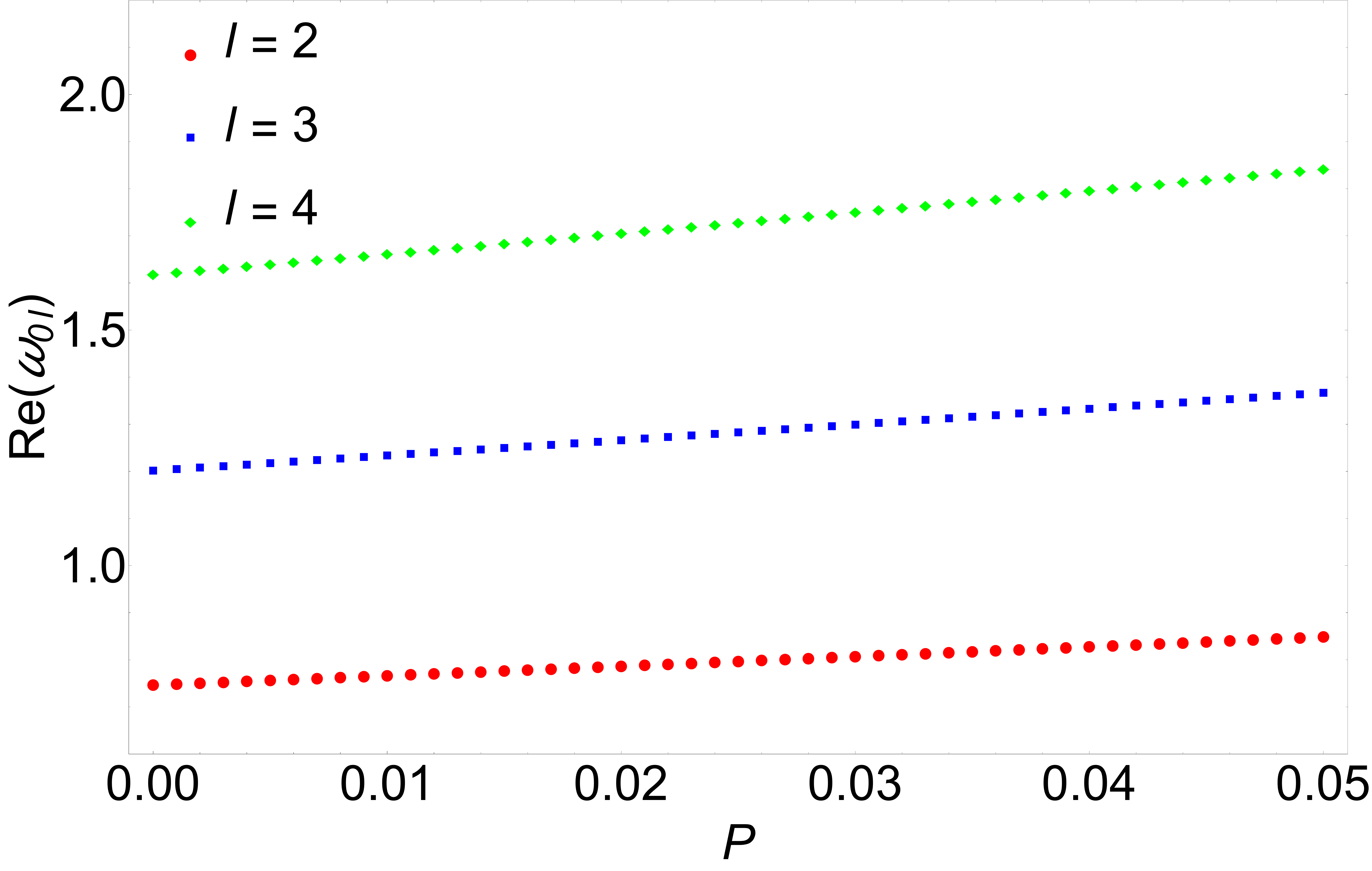}
		\includegraphics[scale =0.14]{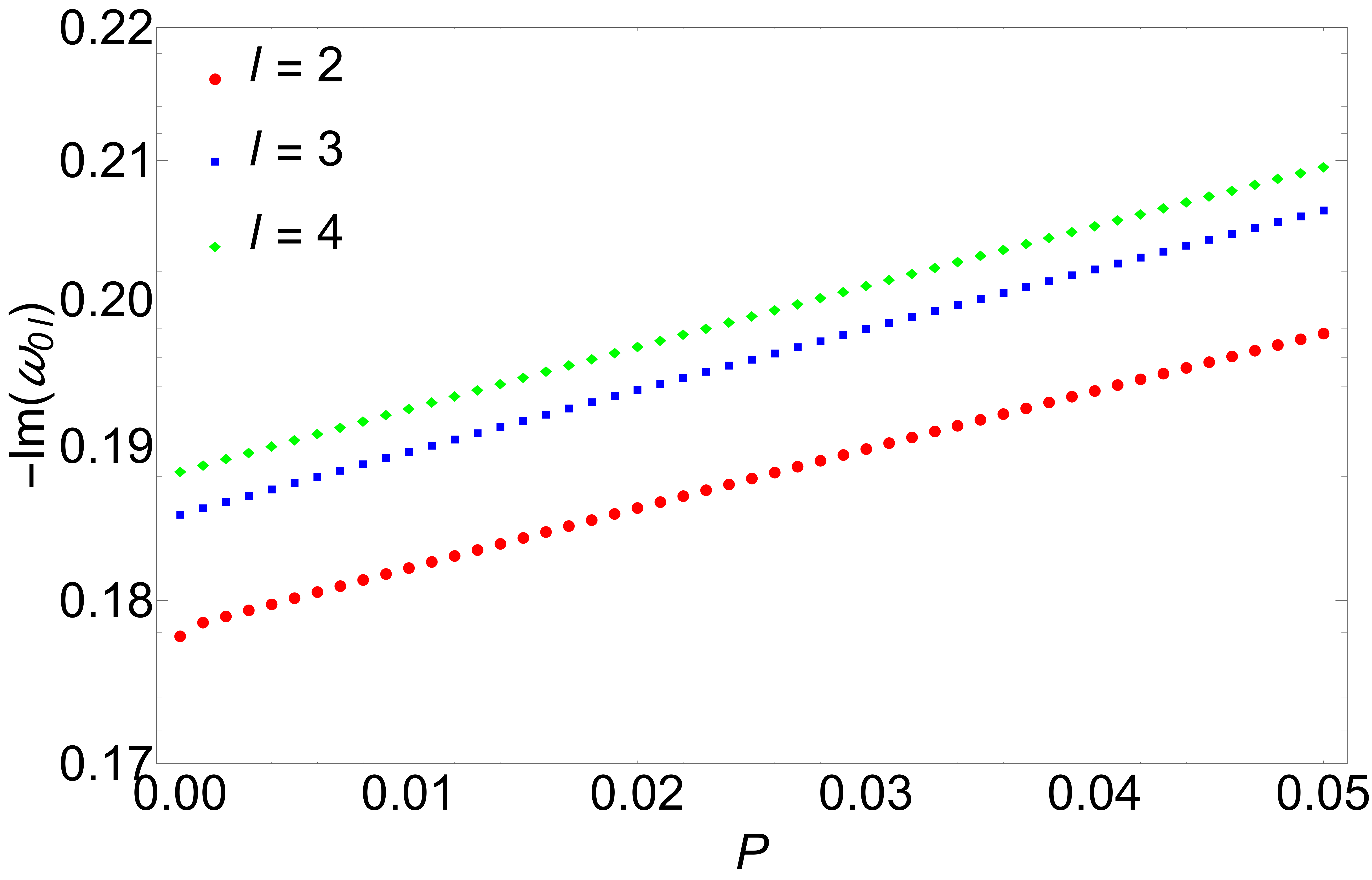}
		\caption{The left plot shows the real parts of $\omega_{02}$, $\omega_{03}$, $\omega_{04}$ in with the parameter $P$. The right plot shows the inverse values of the imaginary  parts of $\omega_{02}$, $\omega_{03}$, $\omega_{04}$ in with the parameter $P$.}
		\label{Re-Im-plot}
	\end{figure}

	In Tables \ref{Tl2}, \ref{Tl3}, and \ref{Tl4}, one can find that when $P=0$, the quasinormal frequencies we obtained agree well with the quasinormal frequencies for the axial gravitational perturbation of the Schwarzschild black hole \cite{Konoplya:2003ii, Cardoso:2003vt}. This is in line with expectations because the metric (\ref{metric}) goes back to the Schwarzschild black hole when both the two quantum parameters vanish. For $l=2,3,4$ and $n=0$, the absolute values of both the real part and the imaginary part of the quasinormal frequencies varying with the value of $P$ are shown in Fig. \ref{Re-Im-plot}. It can be seen that the absolute values of both the real and imaginary parts of the quasinormal frequencies increase with the parameter $P$. This means the parameter $P$ has a positive effect on both the oscillation frequency and decay timescale of the axial gravitational perturbation of the self-dual black hole. This is consistent with the conclusions for the QNMs of the perturbations of the test scalar field and the electromagnetic field of the self-dual black hole \cite{Liu:2020ola, Momennia:2022tug} by taking the ADM mass of the self-dual black hole as a parameter. But it is different from the results in works \cite{Chen:2011zzi, Santos:2021wsw, Cruz:2015bcj, Cruz:2020emz} in which the ADM mass of the self-dual black hole varied during calculation. It is worth mentioning that the effective potential (\ref{V}) we obtain is not the same as the one in \cite{Cruz:2015bcj}, and the effective potential in \cite{Cruz:2015bcj} can not go back to the Schwarzschild case when both the two quantum parameters vanish.
	
	\begin{table*}[t]
		\renewcommand\arraystretch{1.5}
		\centering
		\begin{tabular}{|c|c|c|c|c|c|c|}
			\hline
			\multicolumn{2}{|c|}{$P$} &$ 0 $ & $ 0.001 $ &$ 0.002 $ & $ 0.003 $ & $ 0.004 $ \\
			\hline
			\multirow{2}*{$ \omega_{02} $} & WKB &  $ 0.747239 - 0.177782 i$  & $ 0.749070 - 0.178631 i$ &  $ 0.751042 - 0.179012 i$  & $ 0.753013 - 0.179395 i$ &  $ 0.754989 - 0.179778 i$   \\
			~ &AIM &$ 0.747343 - 0.177925 i$  & $ 0.749308 - 0.178310 i$ &  $ 0.751275 - 0.178696 i$  & $  0.753245 - 0.179083 i$ &  $ 0.755219 - 0.179469 i$   \\
			\hline
			\multirow{2}*{$ \omega_{12} $} & WKB &  $ 0.692593 - 0.546960 i$  & $ 0.694458 - 0.550582 i$ &  $ 0.696397 - 0.551722 i$  & $  0.698308 - 0.552887 i$ &  $ 0.700231 - 0.554047 i$   \\
			~& AIM & $ 0.693422 - 0.547830 i$  & $ 0.695331 - 0.549007 i$ &  $ 0.697244 - 0.550185 i$  & $ 0.699160 - 0.551364 i$ &  $ 0.701079 - 0.552544 i$   \\
			\hline
			\hline
			\multicolumn{2}{|c|}{$P$} &$ 0.005 $ & $ 0.006 $ &$ 0.007 $ & $ 0.008 $ & $ 0.009 $ \\
			\hline
			\multirow{2}*{$ \omega_{02} $} &WKB & $ 0.756968 - 0.180161 i$  & $ 0.758947 - 0.180545i$ &  $ 0.760937 - 0.180927 i$  & $  0.762917 - 0.181313 i$ &  $ 0.764911 - 0.181696 i$   \\
			~ &AIM & $ 0.757195 - 0.179856 i$  & $ 0.759175 - 0.180243 i$ &  $ 0.761158 - 0.180630 i$  & $ 0.763144 - 0.181018 i$ &  $ 0.765133 - 0.181406 i$   \\
			\hline
			\multirow{2}*{$ \omega_{12} $} & WKB & $ 0.702162 - 0.555202 i$  & $ 0.704074 - 0.556377 i$ &  $  0.706039 - 0.557514 i$  & $ 0.707924 - 0.558717 i$ &  $  0.709883 - 0.559865 i$   \\
			~ &AIM  & $ 0.703001 - 0.553724 i$  & $ 0.704927 - 0.554905 i$ &  $ 0.706856 - 0.556086 i$  & $ 0.708788 - 0.557269 i$ &  $ 0.710723 - 0.558451 i$   \\
			\hline
			\hline
			\multicolumn{2}{|c|}{$P$}&$ 0.01 $ & $ 0.02 $ &$ 0.03 $ & $ 0.04 $ & $ 0.05 $ \\
			\hline
			\multirow{2}*{$ \omega_{02} $} & WKB & $ 0.766907 - 0.182080 i$  & $ 0.787023 - 0.185939 i$ &  $ 0.807444 - 0.189824 i$  & $ 0.828184 - 0.193732 i$ &  $ 0.849226 - 0.197664 i$   \\
			~ & AIM & $ 0.767126 - 0.181794 i$  & $0.787220 - 0.185688 i$ &  $ 0.807626 - 0.189606 i$  & $ 0.828342 - 0.193545 i$ &  $ 0.849370 - 0.197505 i$   \\
			\hline
			\multirow{2}*{$ \omega_{12} $} & WKB & $ 0.711840 - 0.561018 i$  & $ 0.731465 - 0.572694 i$ &  $ 0.751382 - 0.584477 i$  & $ 0.771677 - 0.596294 i$ &  $ 0.792250 - 0.608217 i$   \\
			~ & AIM & $  0.712662 - 0.559635 i$  & $ 0.732229 - 0.571507 i$ &  $ 0.752124 - 0.583441 i$  & $0.772348 - 0.595432 i$ &  $  0.792902 - 0.607476 i$   \\
			\hline
		\end{tabular}
		\caption{The QNMs of the axial gravitational perturbation of the self-dual black hole with different values of $P$ and $l = 2~(n<l)$ calculated by the WKB approximation method and the asymptotic iteration method.}
		\label{Tl2}
	\end{table*}

	\begin{table*}[t]
		\renewcommand\arraystretch{1.5}
		\centering
		\begin{tabular}{|c|c|c|c|c|c|c|}
			\hline
			\multicolumn{2}{|c|}{$P$} & $ 0 $ & $0.001$ & $0.002$ & $0.003$ & $0.004$ \\
			\hline
			\multirow{2}*{$ \omega_{03} $} & WKB &  $ 1.198890 - 0.185405 i$  & $ 1.202070 - 0.185814 i$ & $ 1.205260 - 0.186225 i$ & $ 1.208450 - 0.186637 i$ & $ 1.211660 - 0.187049i$ \\
			~ & AIM &  $1.198890 - 0.185406 i$  & $ 1.202070 - 0.185818 i$ & $ 1.205260 - 0.186229 i$ & $ 1.208460 - 0.186641 i$ & $ 1.211660 - 0.187054 i$ \\
			\hline
			\multirow{2}*{$ \omega_{13} $} & WKB &  $ 1.165280 - 0.562581 i$  & $ 1.168370 - 0.563804 i$ & $ 1.171530 - 0.565043 i$ & $ 1.174660 - 0.566297 i$ & $ 1.177840 - 0.567532 i$ \\
			~ & AIM &  $ 1.165290 - 0.562596 i$  & $ 1.168430 - 0.563839 i$ & $ 1.171580 - 0.565083 i$ & $ 1.174740 - 0.566327 i$ & $ 1.177900 - 0.567572 i$ \\
			\hline
			\multirow{2}*{$ \omega_{23} $} & WKB &  $ 1.103190 - 0.958094 i$  & $ 1.105190 - 0.959163 i$ & $ 1.108290 - 0.961239 i$ & $ 1.111290 - 0.963398 i$ & $ 1.114440 - 0.965438i$ \\
			~ & AIM &  $ 1.103370 - 0.958185 i$  & $ 1.106440 - 0.960284 i$ & $ 1.109520 - 0.962384 i$ & $ 1.112600 - 0.964485 i$ & $ 1.115690 - 0.966587i$ \\
			\hline
			\hline
			\multicolumn{2}{|c|}{$P$} & $ 0.005 $ & $0.006$ & $0.007$ & $0.008$ & $0.009$ \\
			\hline
			\multirow{2}*{$ \omega_{03} $} & WKB &  $ 1.214860 - 0.187462 i$  & $ 1.218080 - 0.187874 i$ & $ 1.221290 - 0.188287 i$ & $1.224510 - 0.188700 i$ & $ 1.227740 - 0.189113 i$ \\
			~ & AIM &  $1.214870 - 0.187466 i$  & $ 1.218080 - 0.187879 i$ & $ 1.221300 - 0.188292 i$ & $ 1.224520 - 0.188705 i$ & $ 1.227740 - 0.189119 i$ \\
			\hline
			\multirow{2}*{$ \omega_{13} $} & WKB &  $ 1.180990 - 0.568787 i$  & $ 1.184200 - 0.570014i$ & $ 1.187340 - 0.571279 i$ & $ 1.190530 - 0.572520 i$ & $ 1.193730 - 0.573760 i$ \\
			~ & AIM &  $1.181060 - 0.568818 i$  & $ 1.184230 - 0.570065 i$ & $1.187410 - 0.571312 i$ & $   1.190590 - 0.572560i$ & $ 1.193780 - 0.573809i$ \\
			\hline
			\multirow{2}*{$ \omega_{23} $} & WKB &  $1.117460 - 0.967595 i$  & $ 1.120690 - 0.969581 i$ & $  1.123660 - 0.971795 i$ & $ 1.126810 - 0.973858 i$ & $1.129980 - 0.975910 i$ \\
			~ & AIM &  $1.118780 - 0.968691 i$  & $ 1.121880 - 0.970795i$ & $ 1.124980 - 0.972901 i$ & $ 1.128090 - 0.975008 i$ & $ 1.131200 - 0.977116i$ \\
			\hline
			\hline
			\multicolumn{2}{|c|}{$P$} & $ 0.01 $ & $0.02$ & $0.03$ & $0.04$ & $0.05$ \\
			\hline
			\multirow{2}*{$ \omega_{03} $} & WKB &  $ 1.230970 - 0.189527 i$  & $ 1.263560 - 0.193675 i$ & $ 1.296650 - 0.197847 i$ & $ 1.330240 - 0.202038 i$ & $ 1.364330 - 0.206247 i$ \\
			~ & AIM &  $ 1.230970 - 0.189533 i$  & $ 1.263560 - 0.193683 i$ & $ 1.296660 - 0.197856 i$ & $ 1.330250 - 0.202049 i$ & $ 1.364340 - 0.206260 i$ \\
			\hline
			\multirow{2}*{$ \omega_{13} $} & WKB &  $ 1.196890 - 0.575022i$  & $1.229120 - 0.587533 i$ & $1.261790 - 0.600136 i$ & $ 1.295010 - 0.612780 i$ & $ 1.328740 - 0.625475 i$ \\
			~ & AIM &  $ 1.196970 - 0.575058i$  & $1.229160 - 0.587590 i$ & $ 1.261860 - 0.600186i$ & $ 1.295080 - 0.612838 i$ & $ 1.328790 - 0.625548 i$ \\
			\hline
			\multirow{2}*{$ \omega_{23} $} & WKB &  $1.133000 - 0.978100 i$  & $1.164550 - 0.999115 i$ & $1.196440 - 1.020400 i$ & $1.228970 - 1.041680 i$ & $ 1.262040 - 1.063030 i$ \\
			~ & AIM &  $ 1.134320 - 0.979225 i$  & $1.165780 - 1.000370 i$ & $ 1.197770 - 1.021610 i$ & $ 1.230290 - 1.042940 i$ & $ 1.263430 - 1.064120 i$ \\
			\hline
		\end{tabular}
		\caption{The QNMs of the axial gravitational perturbation of the self-dual black hole with different values of $P$ and $l = 3~(n<l)$ calculated by the WKB approximation method and the asymptotic iteration method.}
		\label{Tl3}
	\end{table*}
	
	\begin{table*}[t]
		\renewcommand\arraystretch{1.5}
		\centering
		\begin{tabular}{|c|c|c|c|c|c|c|}
			\hline
			\multicolumn{2}{|c|}{$P$} & $0$ & $0.001$ & $0.002$ & $0.003$ & $0.004$ \\
			\hline
			\multirow{2}*{$ \omega_{04} $} & WKB & $1.618360 - 0.188328i$ & $1.622670 - 0.188743i$ & $1.626980 - 0.189162i$ & $1.631310 - 0.189581i$ & $1.635640 - 0.190001i$ \\
			~ & AIM & $1.618360 - 0.188328i$ & $1.622670 - 0.188747i$ & $1.626990 - 0.189166i$ & $1.631310 - 0.189586i$ & $1.635640 - 0.190006i$ \\
			\hline
			\multirow{2}*{$ \omega_{14} $} & WKB & $ 1.593260 - 0.568668i$ & $1.597520 - 0.569909i$ & $1.601810 - 0.571169i$ & $1.606090 - 0.572435i$ & $1.610410 - 0.573693i$ \\
			~ & AIM & $1.593260 - 0.568669i$ & $1.597540 - 0.569930i$ & $ 1.601830 - 0.571193i$ & $ 1.606120 - 0.572456i$ & $1.610420 - 0.573720i$ \\
			\hline
			\multirow{2}*{$ \omega_{24} $} & WKB & $1.545390 - 0.959799i$ & $1.549230 - 0.961465i$ & $1.553480 - 0.963572i$ & $1.557680 - 0.965712i$ & $1.561980 - 0.967793i$ \\
			~& AIM & $1.545420 - 0.959816i$ & $1.549640 - 0.961934i$ & $1.553870 - 0.964053i$ & $1.558100 - 0.966173i$ & $1.562340 - 0.968294i$ \\
			\hline
			\multirow{2}*{$ \omega_{34} $} & WKB & $1.479330 - 1.367800i$ & $1.480970 - 1.368580i$ & $1.485160 - 1.371530i$ & $1.489220 - 1.374600i$ & $1.493530 - 1.377460i$ \\
			~& AIM & $1.479670 - 1.367850i$ & $1.483820 - 1.370840i$ & $1.487970 - 1.373830i$ & $1.492120 - 1.376830i$ & $1.496290 - 1.379820i$ \\
			\hline
			\hline
			\multicolumn{2}{|c|}{$P$} & $0.005$ & $0.006$ & $0.007$ & $0.008$ & $0.009$ \\
			\hline
			\multirow{2}*{$ \omega_{04} $} & WKB & $ 1.639980 - 0.190421i$ & $1.644320 - 0.190841i$ & $1.648680 - 0.191261i$ & $ 1.653040 - 0.191682i$ & $1.657400 - 0.192103i$ \\
			~ & AIM & $1.639980 - 0.190426i$ & $1.644320 - 0.190846i$ & $1.648680 - 0.191266i$ & $1.653030 - 0.191687i$ & $1.657400 - 0.192108i$ \\
			\hline
			\multirow{2}*{$ \omega_{14} $} & WKB & $1.614720 - 0.574957i$ & $1.619010 - 0.5762290i$ & $1.623350 - 0.577488i$ & $1.627680 - 0.578753i$ & $1.632010 - 0.580023i$ \\
			~ & AIM & $1.614730 - 0.574985i$ & $1.619040 - 0.576250i$ & $1.623360 - 0.577516i$ & $1.627690 - 0.578783i$ & $1.632030 - 0.580050i$ \\
			\hline
			\multirow{2}*{$ \omega_{24} $} & WKB & $1.566230 - 0.969911i$ & $1.570410 - 0.972080i$ & $1.574750 - 0.974154i$ & $1.579030 - 0.976273i$ & $1.583280 - 0.978414i$ \\
			~ & AIM & $1.566590 - 0.970416i$ & $1.570840 - 0.972540i$ & $1.575110 - 0.974664i$ & $1.579380 - 0.976790i$ & $1.583650 - 0.978917i$ \\
			\hline
			\multirow{2}*{$ \omega_{34} $} & WKB & $1.497710 - 1.380440i$ & $1.501700 - 1.383610i$ & $1.506070 - 1.386430i$ & $1.510280 - 1.389400i$ & $1.514410 - 1.392460i$ \\
			~ & AIM & $1.500460 - 1.382820i$ & $1.504640 - 1.385820i$ & $1.508820 - 1.388820i$ & $1.513200 - 1.391820i$ & $1.517220 - 1.394830i$ \\
			\hline
			\hline
			\multicolumn{2}{|c|}{$P$} & $0.01$ & $0.02$ & $0.03$ & $0.04$ & $0.05$ \\
			\hline
			\multirow{2}*{$ \omega_{04} $} & WKB & $1.661770 - 0.192524i$ & $1.705860 - 0.196748i$ & $1.750630 - 0.200992i$ & $1.796060 - 0.205255i$ & $1.842160 - 0.209535i$ \\
			~ & AIM & $1.661770 - 0.192529i$ & $ 1.705860 - 0.196753i$ & $1.750630 - 0.200998i$ & $1.796060 -0.205262i$ & $1.842160 - 0.209537i$ \\
			\hline
			\multirow{2}*{$ \omega_{14} $} & WKB & $1.636360 - 0.581289i$ & $1.680140 - 0.594005i$ & $1.724610 - 0.606781i$ & $1.769770 - 0.619610 i$ & $1.815590 - 0.632491i$ \\
			~ & AIM & $1.636370 - 0.581318i$ & $ 1.680150 - 0.594035 i$ & $1.724630 - 0.606813i$ & $1.769780 - 0.619645i$ & $1.815590 - 0.632407i$ \\
			\hline
			\multirow{2}*{$ \omega_{24} $} & WKB & $1.587580 - 0.980530i$ & $1.630780 - 1.001860 i$ & $1.674690 - 1.023290i$ & $1.719310 - 1.044790 i$ & $1.764590 - 1.066390i$ \\
			~ & AIM & $1.587940 - 0.981045i$ & $1.631150 - 1.002380i$ & $1.675070 - 1.023810i$ & $1.719690 - 1.045310i$ & $1.765060 - 1.065550i$ \\
			\hline
			\multirow{2}*{$ \omega_{34} $} & WKB & $1.518650 - 1.395430i$ & $1.561030 - 1.425560 i$ & $ 1.604130 - 1.455810i$ & $1.648010 - 1.486120 i$ & $1.692500 - 1.516630i$ \\
			~ & AIM & $1.521430 - 1.397830i$ & $1.563900 - 1.427990i$ & $1.607090 - 1.458230i$ & $1.651150 - 1.488830i$ & $1.699090 - 1.510890i$ \\
			\hline
		\end{tabular}
		\caption{The QNMs of the axial gravitational perturbation of the self-dual black hole with different values of $P$ and $l = 4~(n<l)$ calculated by the WKB approximation method and the asymptotic iteration method.}
		\label{Tl4}
	\end{table*}
	
	\subsection{Ringdown Waveforms}
\begin{figure}[h]\centering
		\includegraphics[scale =0.15]{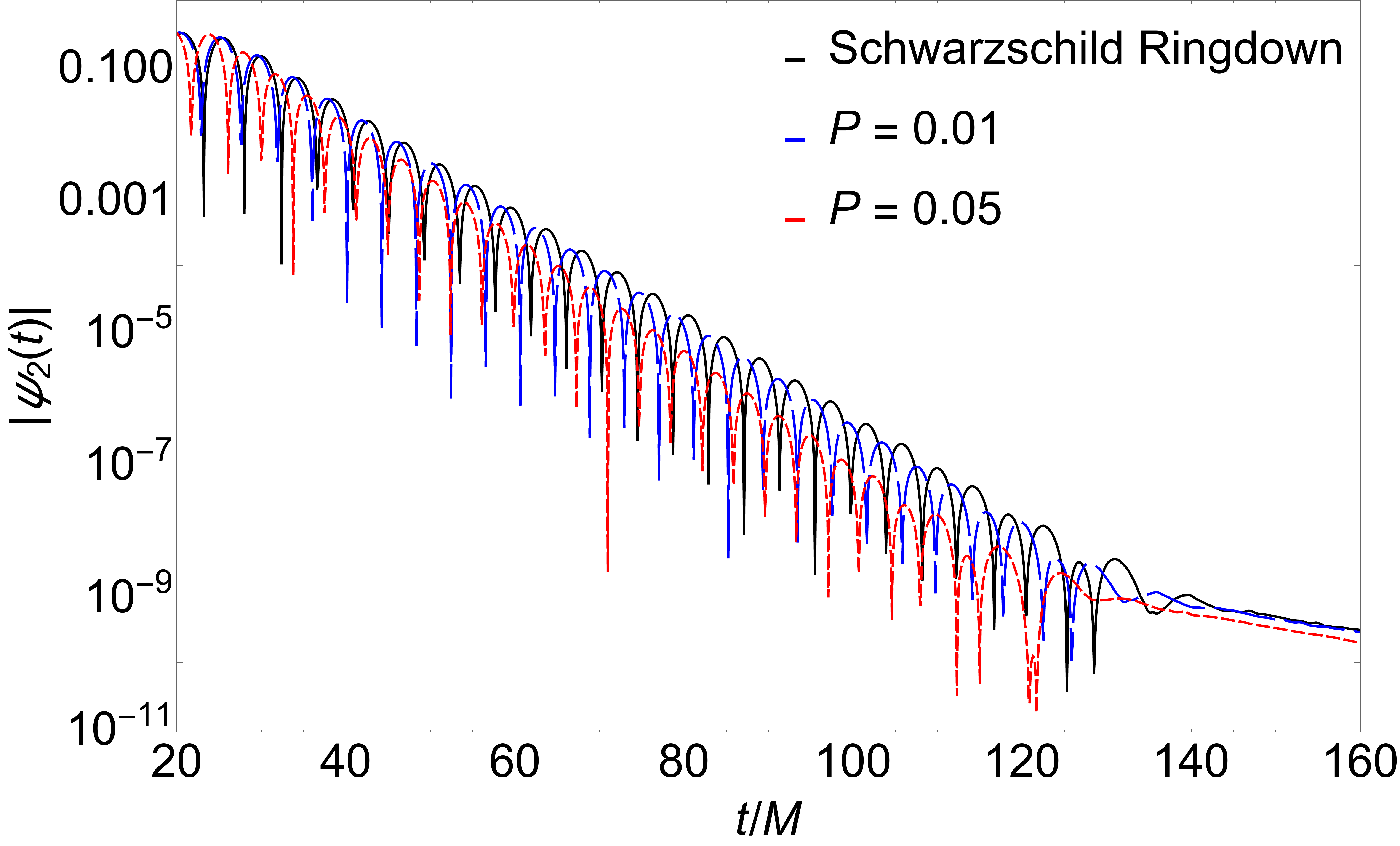}
		\caption{The time evolution of the wave function $\Psi_2(t)$ ($l=2$) of the axial gravitational perturbation for the self-dual black hole with different values of the parameter $P$, evaluated at $r = 10 r_{+}$. The black curve ($P = 0$) shows the Schwarzschild ringdown case.}
		\label{ringdown-plot}
	\end{figure}

	To investigate the contribution of all modes of the axial perturbation of the self-dual black hole, we can consider the numeric evolution of an initial wave packet in the self-dual black hole spacetime. In a finite time domain, the Schr\"{o}dinger-like equation (\ref{RW-eq}) can be rewritten as
	\bqn\lb{-eq}
	\frac{\p^2 \Psi}{\p r^2_{\ast}} - \f{\p^2 \Psi}{\p t^2} - V (r_\ast) \Psi =0.
	\eqn
	Using the light-cone coordinates $u = t - r_{\ast}$ and $ v = t + r_{\ast}$ \cite{Gundlach:1993tp}, the above equation can be written as
	\bqn\lb{ringdown-eq}
	4 \frac{\p ^2 \Psi (u, v)}{\p u \p v} - V (u, v) \Psi (u,v) = 0.
	\eqn
	Here, we set the initial data for Eq.~(\ref{ringdown-eq}) as
	\bqn
	\Psi (u, 0) = 0~~\text{and}~~\Psi (0,v) = \tx{exp} \left( - \frac{(v-v_{c})^2}{2 \beta^2} \right),
	\eqn
where $\Psi (0,v)$ is a Gaussian wave packet centered in $v_c$ and having width $\beta$. Then, we choose the observer located at $r = 10 r_{+}$ and numerically solve the partial differential equation (\ref{ringdown-eq}) to generate the ringdown waveforms. As shown in Fig.~\ref{ringdown-plot}, the waveform with a larger value of the parameter $P$ damps more quickly. Finally, without loss of generality, we use a modified exponentially decaying function $ e^{\omega_I t} A \sin(\omega_R + B)$ to fit the data in Fig.~\ref{ringdown-plot} and calculate the fundamental mode $\omega_{02}$ with different values of the parameter $P$, which plays a major role in the ringdown waveforms. The results are shown in Tab.~\ref{ringdown-T}. Considering the error in the numerical calculation process, one can find that the fitting values of the fundamental mode $\omega_{02}$  with different values of the parameter $P$ in Tab.~\ref{ringdown-T} agree well with the results obtained by using the WKB approximation method and the asymptotic iteration method.
	
	\begin{table*}[t]
		\renewcommand\arraystretch{1.5}
		\centering
		\begin{tabular}{|c|c|c|c|c|}
			\hline
			\multicolumn{2}{|c|}{$P$} & $0$  & $0.01$  & $0.05$  \\
			\hline
			\multirow{3}*{$ \omega_{02} $}   & Fitting & $ 0.747304 - 0.178066 i$ & $ 0.765487 - 0.182486 i$ & $ 0.844821 - 0.199696 i$ \\
			~& WKB & $ 0.747239 - 0.177782 i$ & $ 0.766907 - 0.182080 i$ & $ 0.849226 - 0.197664 i$ \\
			~& AIM & $ 0.747343 - 0.177925 i$ & $ 0.767126 - 0.181794 i$ & $ 0.849370 - 0.197505 i$ \\
			\hline
		\end{tabular}
		\caption{The fundamental mode $\omega_{02}$ calculated by fitting the data in Fig.~\ref{ringdown-plot}, WKB approximation method, and the asymptotic iteration method.}
		\label{ringdown-T}
	\end{table*}

	\section {QNMs in the eikonal limit and circular null geodesics}\lb{sec4}
	
	Assuming a stationary, spherically symmetric, and asymptotically flat line element, Cardoso $et~al.$ showed that, in the eikonal limit $l\rightarrow \infty$, the QNMs of a black hole in any dimensions are \cite{Cardoso:2008bp}
	\bqn\lb{relation-general}
	\omega_{\text{QNM}} = \Omega_c l - i (n+1/2) |\lambda_c|,
	\eqn
	where the subscript $c$ means that the quantity is evaluated at the radius $r = r_c$ of a circular null geodesic, $\Omega_c$ and $\lambda_c$ are the coordinate angular velocity and the Lyapunov exponent of the circular null geodesics, respectively.
	It is an interesting relation between quasinormal frequencies and the parameters of the circular null geodesics, but it may be not valid in a specific black hole \cite{Konoplya:2017wot}. In this section, based on Eq.~(\ref{relation-general}), we shall derive the explicit relation between the QNMs, in the eikonal limit, and the parameters of the circular null geodesics of the self-dual black hole. Then we numerically verify this relation.

The Lagrangian for a photon in the equatorial plane ($\theta=\pi/2$) in the self-dual black hole is
	\bqn\lb{Largrangian}
	\mathscr{L} = \frac{1}{2} \left[ -f(r) \dot{t}^2 + \frac{1}{g(r)} \dot{r}^2 + h(r) \dot{\varphi}^2 \right],
	\eqn
	and the generalized momentum from this Lagrangian is
	\bqn
	p_{t} &=& - f(r) \dot{t} = -E, \lb{E} \\
	p_{\varphi} &=& h(r) \dot{\varphi} = L, \lb{L} \\
	p_{r} &=& \frac{\dot{r}}{g(r)} ,
	\eqn
	where $E$ is the energy, $L$ is the angular momentum, and the dot denotes differentiation to an affine parameter along the geodesics of the photon. Because the Lagrangian (\ref{Largrangian}) is independent of both $t$ and $\varphi$, $E$ and $L$ are conserved. From Eqs. (\ref{E}) and (\ref{L}), one can get
	\bqn \lb{dphidt}
	\dot{\varphi} = \frac{L}{h(r)},~~ \dot{t} = \frac{E}{f(r)}.
	\eqn
	The Hamiltonian for the photon is
	\bqn\lb{Hamiltonian}
	\mathscr{H}
	&=& \frac{1}{2} \left[- E \dot{t} + L \dot{\varphi} + \frac{1}{g(r)} \dot{r}^2 \right] = 0.
	\eqn
	With Eqs. (\ref{dphidt}) and (\ref{Hamiltonian}), one can define the effective potential as
	\bqn\lb{popential}
	V_{r} \equiv \dot{r}^2 = g(r) \left[ \frac{E^2}{f(r)} - \frac{L^2}{h(r)} \right].
	\eqn
	For the circular ($r=r_{c}$) null geodesics on the equatorial plane, the conditions $V_r = V_r'=0$ lead to
	\bqn
	\frac{E}{L} = \pm \sqrt{\frac{f_c}{h_c}},~~f_c h_c' = f_c' h_c.
	\eqn
	In this case
	\bqn
	V_r''= \frac{L^2 g_c}{f_c h_c^2} \left( f_c h_c'' - f_c'' h_c \right),
	\eqn
	and the coordinate angular velocity is
	\bqn\lb{Omega_c}
	\Omega_{c} = \frac{\dot{\varphi}}{\dot{t}} = \left( \frac{f_c}{h_c} \right)^{1/2}.
	\eqn
	The principal Lyapunov exponent is a quantity that characterizes the rate of separation of infinitesimally close geodesics. Using the expression of the principal Lyapunov exponent \cite{Cardoso:2008bp}
	\bqn
	\lambda = \sqrt{\frac{V_r''}{2 \dot{t}^2}},
	\eqn
	we get
	\bqn\lb{lambda_c}
	\lambda_c = \sqrt{\frac{g_c}{2 h_c} \left(f_c h_c'' - f_c'' h_c \right)}
	\eqn
	for the circular null geodesics of the self-dual black hole.~Taking Eqs.~(\ref{Omega_c}) and (\ref{lambda_c}) into (\ref{relation-general}), we derive the relation between the QNMs, in the eikonal limit, and the parameters of the circular null geodesics of the self-dual black hole
	\bqn\lb{relation-SDBH}
	\omega_{\text{QNM}} = l \left( \frac{f_c}{h_c} \right)^{1/2} - i (n+1/2) \sqrt{\frac{g_c}{2 h_c} \left(f_c h_c'' - f_c'' h_c \right)}.
	\eqn
	Setting $l=100$ and ${n = 0, 1, 2, 3, 4}$, we calculate the quasinormal frequencies in the eiknoal limit by the WKB approximation method and the relation (\ref{relation-SDBH}). We list the numerical results in Tab.~\ref{Tl100}. Considering the error of calculation, one can find that the quasinormal frequencies obtained by the WKB approximation method and the relation (\ref{relation-SDBH}) agree well with each other in Tab.~\ref{Tl100}. It means that the general relation (\ref{relation-general}), between quasinormal frequencies and the parameters of the circular null geodesics, is right for the self-dual black hole in loop quantum gravity.
	
	\begin{table*}[t]
		\renewcommand\arraystretch{1.5}
		\centering
		\begin{tabular}{|c|c|c|c|c|c|c|}
			\hline
			\multicolumn{2}{|c|}{$P$} & $0$ & $0.001$ & $0.002$ & $0.003$ & $0.004$ \\
			\hline
			\multirow{2}*{$ \omega_{0,100} $} & WKB & $ 38.6775 - 0.19244 i$ & $ 38.7807 - 0.19287i$ & $ 38.8841 - 0.19330i$ & $38.9876 - 0.19373 i$ & $39.0913 - 0.19416 i$ \\
			~ & CNG & $38.4900 - 0.19245 i$ & $38.5927 - 0.19288 i$ & $38.6956 - 0.19331 i$ & $38.7987 - 0.19373 i$ & $38.9019 - 0.19416 i$ \\
			\hline
			\multirow{2}*{$ \omega_{1,100} $} & WKB & $ 38.6764 - 0.57733i$ & $38.7796 - 0.57862 i$ & $ 38.8830 - 0.57990i$ & $ 38.9866 - 0.58119i$ & $ 39.0903 - 0.58247i$ \\
			~ & CNG & $38.4900 - 0.57735 i$ & $38.5927 - 0.57863 i$ & $38.6956 - 0.57992 i$ & $ 38.7987 - 0.58120 i$ & $ 38.9019 - 0.58249i$ \\
			\hline
			\multirow{2}*{$ \omega_{2,100} $} & WKB & $ 38.6743 - 0.96225 i$ & $38.7775 - 0.96439 i$ & $ 38.8809 - 0.96653i$ & $ 38.9844 - 0.96867i$ & $39.0881 - 0.97081 i$ \\
			~& CNG & $38.4900 - 0.96225 i$ & $38.5927 - 0.96439 i$ & $38.6956 - 0.96653 i$ & $38.7987 - 0.96867 i$ & $ 38.9019 - 0.97081i$ \\
			\hline
			\multirow{2}*{$ \omega_{3,100} $} & WKB & $ 38.6711 - 1.34719 i$ & $ 38.7743 - 1.35019i$ & $ 38.8777 - 1.35318i$ & $38.9812 - 1.35618 i$ & $39.0849 - 1.35918 i$ \\
			~& CNG & $38.4900 - 1.34715 i$ & $38.5927 - 1.35015 i$ & $38.6956 - 1.35314 i$ & $38.7987 - 1.35614 i$ & $38.9019 - 1.35914 i$ \\
			\hline
			\multirow{2}*{$ \omega_{4,100} $} & WKB & $ 38.6668 - 1.73219 i$ & $ 38.7701 - 1.73603i$ & $ 38.8734 - 1.73989i$ & $ 38.9770 - 1.74374i$ & $ 39.0806 - 1.74760i$ \\
			~& CNG & $38.4900 - 1.73205 i$ & $38.5927 - 1.73590 i$ & $38.6956 - 1.73975 i$ & $38.7987 - 1.74361 i$ & $ 38.9019 - 1.74746i$ \\
			\hline
			\hline
			\multicolumn{2}{|c|}{$P$} & $0.005$ & $0.006$ & $0.007$ & $0.008$ & $0.009$ \\
			\hline
			\multirow{2}*{$ \omega_{0,100} $} & WKB & $39.1952 - 0.19458 i$ & $39.2992 - 0.19501 i$ & $ 39.4034 - 0.19544i$ & $ 39.5077 - 0.19587i$ & $39.6122 - 0.19630 i$ \\
			~ & CNG & $ 39.0052 - 0.19459i$ & $ 39.1087 - 0.19502i$ & $39.2124 - 0.19545 i$ & $39.3162 - 0.19588 i$ & $39.4202 - 0.19631 i$ \\
			\hline
			\multirow{2}*{$ \omega_{1,100} $} & WKB & $39.1941 - 0.58376 i$ & $39.2981 - 0.58504 i$ & $ 39.4023 - 0.58633i$ & $39.5067 - 0.58762 i$ & $ 39.6112 - 0.58891i$ \\
			~ & CNG & $ 39.0052 - 0.58377i$ & $ 39.1087 - 0.58506i$ & $ 39.2124 - 0.58635i$ & $39.3162 - 0.58764 i$ & $ 39.4202 - 0.58892i$ \\
			\hline
			\multirow{2}*{$ \omega_{2,100} $} & WKB & $39.1920 - 0.97295 i$ & $ 39.2960 - 0.97510i$ & $ 39.4002 - 0.97724i$ & $ 39.5045 - 0.97939i$ & $39.6090 - 0.98154 i$ \\
			~& CNG & $39.0052 - 0.97296 i$ & $ 39.1087 - 0.97510i$ & $ 39.2124 - 0.97725i$ & $ 39.3162 - 0.97939i$ & $ 39.4202 - 0.98154i$ \\
			\hline
			\multirow{2}*{$ \omega_{3,100} $} & WKB & $39.1888 - 1.36218 i$ & $39.2928 - 1.36518 i$ & $ 39.3970 - 1.36819i$ & $39.5013 - 1.37119 i$ & $39.6058 - 1.37420 i$ \\
			~& CNG & $39.0052 - 1.36214 i$ & $39.1087 - 1.36514 i$ & $39.2124 - 1.36814 i$ & $39.3162 - 1.37115 i$ & $39.4202 - 1.37416 i$ \\
			\hline
			\multirow{2}*{$ \omega_{4,100} $} & WKB & $ 39.1845 - 1.75146i$ & $ 39.2885 - 1.75532i$ & $ 39.3927 - 1.75918i$ & $ 39.4970 - 1.76305i$ & $39.6015 - 1.76691 i$ \\
			~& CNG & $39.0052 - 1.75132 i$ & $39.1087 - 1.75518 i$ & $39.2124 - 1.75904 i$ & $39.3162 - 1.76291 i$ & $ 39.4202 - 1.76677i$ \\
			\hline
			\hline
			\multicolumn{2}{|c|}{$P$} & $0.01$ & $0.02$ & $0.03$ & $0.04$ & $0.05$ \\
			\hline
			\multirow{2}*{$ \omega_{0,100} $} & WKB & $ 39.7169 - 0.19673i$ & $ 40.7723 - 0.20104 i$ & $ 41.8436 - 0.20537 i$ & $ 42.9308 - 0.20971 i$ & $ 44.0338 - 0.21406i$ \\
			~ & CNG & $39.5244 - 0.19674 i$ & $ 40.5747 - 0.20105 i$ & $ 41.6408 - 0.20537 i$ & $42.7227 -0.20972 i$ & $ 43.8204 - 0.21407i$ \\
			\hline
			\multirow{2}*{$ \omega_{1,100} $} & WKB & $39.7158 - 0.59020 i$ & $40.7712 - 0.60312 i$ & $ 41.8425 - 0.61610 i$ & $42.9297 - 0.62913 i$ & $ 44.0327 - 0.64220i$ \\
			~ & CNG & $39.5244 - 0.59021 i$ & $ 40.5747 - 0.60314 i$ & $ 41.6408 - 0.61612 i$ & $42.7227 -0.62915 i$ & $ 43.8204 - 0.64222i$ \\
			\hline
			\multirow{2}*{$ \omega_{2,100} $} & WKB & $39.7137 - 0.98368 i$ & $40.7690 - 1.00523  i$ & $ 41.8403 - 1.02686 i$ & $42.9275 - 1.04857 i$ & $44.0304 - 1.07036 i$ \\
			~& CNG & $39.5244 - 0.98369 i$ & $ 40.5747 - 1.00523 i$ & $ 41.6408 - 1.02686 i$ & $42.7227 -1.04858 i$ & $43.8204 - 1.07036 i$ \\
			\hline
			\multirow{2}*{$ \omega_{3,100} $} & WKB & $ 39.7104 - 1.37721i$ & $ 40.7658 - 1.40737 i$ & $ 41.8370 - 1.43765 i$ & $ 42.9241 - 1.46805i$ & $44.0270 - 1.49855 i$ \\
			~& CNG & $39.5244 - 1.37716 i$ & $ 40.5747 - 1.40732 i$ & $ 41.6408 - 1.43761 i$ & $42.7227 -1.46801 i$ & $43.8204 - 1.49851 i$ \\
			\hline
			\multirow{2}*{$ \omega_{4,100} $} & WKB & $ 39.7061 - 1.77078i$ & $ 40.7614 - 1.80955 i$ & $ 41.8326 - 1.84850 i$ & $ 42.9197 - 1.88758i$ & $ 44.0225 - 1.92680 i$ \\
			~& CNG & $ 39.5244 - 1.77064i$ & $ 40.5747 - 1.80941 i$ & $ 41.6408 - 1.84836 i$ & $42.7227 -1.88744 i$ & $43.8204 - 1.92665 i$ \\
			\hline
		\end{tabular}
		\caption{The QNMs $\omega_{nl}$ of the axial gravitational perturbation of the self-dual black hole with different values of the parameter $P$ and $l = 100 $,  calculated by the WKB approximation method and the QNMs-circular null geodesics (CNG) relation (\ref{relation-SDBH}).}
		\label{Tl100}
	\end{table*}
	
	\section{Conclusions and Discussions}\lb{sec5}
	
	In this work, we investigated the QNMs of the axial gravitational perturbation of the self-dual black hole with the fixed ADM mass in loop quantum gravity. Simulating the quantum correction by an effective anisotropic matter fluid, we obtained the master equation of the axial gravitational perturbation of the self-dual black hole. We considered the influence of the quantum parameter $P$, and found that the height of the effective potential increases with the parameter $P$. Using the WKB approximation method and the asymptotic iteration method, we calculated the QNMs of the axial gravitational perturbation of the self-dual black hole with different values of the parameter $P$. We found that the parameter $P$ has a positive effect on the absolute values of both the real part and imaginary part of the quasinormal frequency. This result is consistent with the conclusions for the QNMs of the perturbation of the scalar field and the electromagnetic field on the self-dual black hole with the fixed ADM mass~\cite{Liu:2020ola, Momennia:2022tug}. In the eikonal limit, we obtained the relation between the QNMs and the parameters of the circular null geodesics in the self-dual black hole, and numerically verify it.

With more and more gravitational wave signals of compact binary components detected by the LIGO-Virgo-KAGRA collaboration and pictures of supermassive objects taken by the Event Horizon Telescope, it is possible to test gravitational theories in the strong gravitational field with multi-messenger.  And the coupling of electromagnetic and gravitational fields should be considered \cite{Zou:2021lkj, Guo:2022rms}. Cardoso $et~al.$ provided the parameterized black hole quasinormal ringdown in the general spherical symmetry background spacetime \cite{McManus:2019ulj, Cardoso:2019mqo}. V\"olkel $et~al.$ got the bounds on modifications of black hole perturbation potentials near the light ring \cite{Volkel:2022khh}. The shadow and ring of a black hole may exhibit rich behavior \cite{Feng:2020tyc, Yifu, Zeng:2021mok, Chen:2022qrw, Liu:2022ruc}, and the relation between QNMs and shadow should be further explored. It would be interesting to find the constraints on the quantum correction parameters of the black holes in loop quantum gravity with the observed gravitational wave ringdown signals and pictures of black holes. And thermodynamics is also a fundamental property of black holes \cite{Wei:2021bwy, Cai:2021sag, Song:2020arr}, the relation between QNMs and the thermodynamics of a black hole  deserves attention. On the other hand, black holes always rotate in the real world, so the gravitational perturbations of the rotating black holes in loop quantum gravity and the related properties should be considered in future work.

\section*{Acknowledgements}

We thank Tao Zhu for important suggestion. This work was supported by National Key Research and Development Program of China (Grant No. 2020YFC2201503), the National Natural Science Foundation of China (Grants No. 12205129, No. 12147166, No. 11875151, No. 12075103, and No. 12247101), the China Postdoctoral Science Foundation (Grant No. 2021M701529), the 111 Project (Grant No. B20063), and Lanzhou City’s scientific research funding subsidy to Lanzhou University.



\begin{thebibliography}{369}
	
\bibitem{LIGOScientific:2016aoc}
B.~P.~Abbott \textit{et al.} [LIGO Scientific and Virgo], Observation of Gravitational Waves from a Binary Black Hole Merger,
\href{\doibase 10.1103/PhysRevLett.116.061102}{Phys. Rev. Lett. \textbf{116}, 061102 (2016)}, [arXiv:1602.03837 [gr-qc]].

\bibitem{Cai:2017cbj}
R.~G.~Cai, Z.~Cao, Z.~K.~Guo, S.~J.~Wang, and T.~Yang, The Gravitational-Wave Physics,
\href{\doibase doi:10.1093/nsr/nwx029}{Natl. Sci. Rev. \textbf{4} (2017) no.5, 687-706},
[arXiv:1703.00187 [gr-qc]].

\bibitem{Bian:2021ini}
L.~Bian, R.~G.~Cai, S.~Cao, Z.~Cao, H.~Gao, Z.~K.~Guo, K.~Lee, D.~Li, J.~Liu, and Y.~Lu, \textit{et al.}
The Gravitational-wave physics II: Progress,
\href{\doibase doi:10.1007/s11433-021-1781-x}{Sci. China Phys. Mech. Astron. \textbf{64} (2021) no.12, 120401},
[arXiv:2106.10235 [gr-qc]].
	
\bibitem{EventHorizonTelescope:2019dse}
K.~Akiyama \textit{et al.} [Event Horizon Telescope],
First M87 Event Horizon Telescope Results. I. The Shadow of the Supermassive Black Hole,
\href{\doibase 10.3847/2041-8213/ab0ec7}{Astrophys. J. Lett. \textbf{875}, L1 (2019)},
[arXiv:1906.11238 [astro-ph.GA]].
	
\bibitem{EventHorizonTelescope:2019uob}
K.~Akiyama \textit{et al.} [Event Horizon Telescope],
First M87 Event Horizon Telescope Results. II. Array and Instrumentation,
\href{\doibase 10.3847/2041-8213/ab0c96}{Astrophys. J. Lett. \textbf{875}, L2 (2019)},
[arXiv:1906.11239 [astro-ph.IM]].
	
\bibitem{EventHorizonTelescope:2019jan}
K.~Akiyama \textit{et al.} [Event Horizon Telescope],
First M87 Event Horizon Telescope Results. III. Data Processing and Calibration,
\href{\doibase 10.3847/2041-8213/ab0c57}{Astrophys. J. Lett. \textbf{875}, L3 (2019)},
[arXiv:1906.11240 [astro-ph.GA]].
	
\bibitem{EventHorizonTelescope:2019ths}
K.~Akiyama \textit{et al.} [Event Horizon Telescope],
First M87 Event Horizon Telescope Results. IV. Imaging the Central Supermassive Black Hole,
\href{\doibase 10.3847/2041-8213/ab0e85}{Astrophys. J. Lett. \textbf{875}, L4 (2019)},
[arXiv:1906.11241 [astro-ph.GA]].
	
\bibitem{EventHorizonTelescope:2019pgp}
K.~Akiyama \textit{et al.} [Event Horizon Telescope],
First M87 Event Horizon Telescope Results. V. Physical Origin of the Asymmetric Ring,
\href{\doibase 10.3847/2041-8213/ab0f43}{Astrophys. J. Lett. \textbf{875}, L5 (2019)},
[arXiv:1906.11242 [astro-ph.GA]].
	
\bibitem{EventHorizonTelescope:2019ggy}
K.~Akiyama \textit{et al.} [Event Horizon Telescope],
First M87 Event Horizon Telescope Results. VI. The Shadow and Mass of the Central Black Hole,
\href{\doibase 10.3847/2041-8213/ab1141}{Astrophys. J. Lett. \textbf{875}, L6 (2019)},
[arXiv:1906.11243 [astro-ph.GA]].
	
\bibitem{EventHorizonTelescope:2022wkp}
K.~Akiyama \textit{et al.} [Event Horizon Telescope],
First Sagittarius A* Event Horizon Telescope Results. I. The Shadow of the Supermassive Black Hole in the Center of the Milky Way,
\href{\doibase 10.3847/2041-8213/ac6674}{Astrophys. J. Lett. \textbf{930}, L12 (2022)}.
	
\bibitem{EventHorizonTelescope:2022apq}
K.~Akiyama \textit{et al.} [Event Horizon Telescope],
First Sagittarius A* Event Horizon Telescope Results. II. EHT and Multiwavelength Observations, Data Processing, and Calibration,
\href{\doibase 10.3847/2041-8213/ac6675}{Astrophys. J. Lett. \textbf{930}, L13 (2022)}.
	
\bibitem{EventHorizonTelescope:2022wok}
K.~Akiyama \textit{et al.} [Event Horizon Telescope],
First Sagittarius A* Event Horizon Telescope Results. III. Imaging of the Galactic Center Supermassive Black Hole,
\href{\doibase 10.3847/2041-8213/ac6429}{Astrophys. J. Lett. \textbf{930}, L14 (2022)}.
	
\bibitem{EventHorizonTelescope:2022exc}
K.~Akiyama \textit{et al.} [Event Horizon Telescope],
First Sagittarius A* Event Horizon Telescope Results. IV. Variability, Morphology, and Black Hole Mass,
\href{\doibase 10.3847/2041-8213/ac6736}{Astrophys. J. Lett. \textbf{930}, L15 (2022)}.
	
\bibitem{EventHorizonTelescope:2022urf}
K.~Akiyama \textit{et al.} [Event Horizon Telescope],
First Sagittarius A* Event Horizon Telescope Results. V. Testing Astrophysical Models of the Galactic Center Black Hole,
\href{\doibase 10.3847/2041-8213/ac6672}{Astrophys. J. Lett. \textbf{930}, L16 (2022)}.
	
\bibitem{EventHorizonTelescope:2022xqj}
K.~Akiyama \textit{et al.} [Event Horizon Telescope],
First Sagittarius A* Event Horizon Telescope Results. VI. Testing the Black Hole Metric,
\href{\doibase 10.3847/2041-8213/ac6756}{Astrophys. J. Lett. \textbf{930}, L17 (2022)}.
	
\bibitem{LIGOScientific:2018mvr}
B.~P.~Abbott \textit{et al.} [LIGO Scientific and Virgo],
GWTC-1: A Gravitational-Wave Transient Catalog of Compact Binary Mergers Observed by LIGO and Virgo during the First and Second Observing Runs,
\href{\doibase 10.1103/PhysRevX.9.031040}{Phys. Rev. X \textbf{9}, 031040 (2019)},
[arXiv:1811.12907 [astro-ph.HE]].
	
\bibitem{LIGOScientific:2020ibl}
R.~Abbott \textit{et al.} [LIGO Scientific and Virgo],
GWTC-2: Compact Binary Coalescences Observed by LIGO and Virgo During the First Half of the Third Observing Run,
\href{\doibase 10.1103/PhysRevX.11.021053}{Phys. Rev. X \textbf{11}, 021053 (2021)},
[arXiv:2010.14527 [gr-qc]].
	
\bibitem{LIGOScientific:2021usb}
R.~Abbott \textit{et al.} [LIGO Scientific and VIRGO],
GWTC-2.1: Deep Extended Catalog of Compact Binary Coalescences Observed by LIGO and Virgo During the First Half of the Third Observing Run,
\href{https://arxiv.org/abs/2108.01045}{[arXiv:2108.01045 [gr-qc]]}.
	
\bibitem{LIGOScientific:2021djp}
R.~Abbott \textit{et al.} [LIGO Scientific, VIRGO and KAGRA],
GWTC-3: Compact Binary Coalescences Observed by LIGO and Virgo During the Second Part of the Third Observing Run',
\href{https://arxiv.org/abs/2111.03606}{[arXiv:2111.03606 [gr-qc]]}.
	
\bibitem{LIGOScientific:2016lio}
B.~P.~Abbott \textit{et al.} [LIGO Scientific and Virgo],
Tests of general relativity with GW150914,
\href{\doibase doi:10.1103/PhysRevLett.116.221101}{Phys. Rev. Lett. \textbf{116}, 221101 (2016)},
[erratum: Phys. Rev. Lett. \textbf{121} (2018), 129902]
[arXiv:1602.03841 [gr-qc]].
	
\bibitem{LIGOScientific:2019fpa}
B.~P.~Abbott \textit{et al.} [LIGO Scientific and Virgo],
Tests of General Relativity with the Binary Black Hole Signals from the LIGO-Virgo Catalog GWTC-1,
\href{\doibase doi:10.1103/PhysRevD.100.104036}{Phys. Rev. D \textbf{100}, 104036 (2019)},
[arXiv:1903.04467 [gr-qc]].
	
\bibitem{LIGOScientific:2020tif}
R.~Abbott \textit{et al.} [LIGO Scientific and Virgo],
Tests of general relativity with binary black holes from the second LIGO-Virgo gravitational-wave transient catalog,
\href{\doibase 10.1103/PhysRevD.103.122002}{Phys. Rev. D \textbf{103}, 122002 (2021)},
[arXiv:2010.14529 [gr-qc]].
	
\bibitem{LIGOScientific:2021sio}
R.~Abbott \textit{et al.} [LIGO Scientific, VIRGO and KAGRA],
Tests of General Relativity with GWTC-3,
\href{https://arxiv.org/abs/2112.06861}{[arXiv:2112.06861 [gr-qc]]}.
	
\bibitem{Chandrasekhar:1985kt}
S.~Chandrasekhar, The mathematical theory of black holes, Oxford University Press, New York, 1983.
	
\bibitem{Maggiore:2018sht}
M.~Maggiore,
Gravitational Waves. Vol. 2: Astrophysics and Cosmology,
Oxford University Press, 2018,
ISBN 978-0-19-857089-9.
	
\bibitem{Kokkotas:1999bd}
K.~D.~Kokkotas and B.~G.~Schmidt,
Quasinormal modes of stars and black holes,
\href{\doibase 10.12942/lrr-1999-2}{Living Rev. Rel. \textbf{2}, 2 (1999)},
[arXiv:gr-qc/9909058 [gr-qc]].
	
\bibitem{Nollert:1999ji}
H.~P.~Nollert,
TOPICAL REVIEW: Quasinormal modes: the characteristic `sound' of black holes and neutron stars,
\href{\doibase 10.1088/0264-9381/16/12/201}{Class. Quant. Grav. \textbf{16}, R159-R216 (1999)}.
	
\bibitem{Berti:2009kk}
E.~Berti, V.~Cardoso, and A.~O.~Starinets,
Quasinormal modes of black holes and black branes,
\href{\doibase 10.1088/0264-9381/26/16/163001}{Class. Quant. Grav. \textbf{26}, 163001 (2009)},
[arXiv:0905.2975 [gr-qc]].
	
\bibitem{Konoplya:2011qq}
R.~A.~Konoplya and A.~Zhidenko,
Quasinormal modes of black holes: From astrophysics to string theory,
\href{\doibase 10.1103/RevModPhys.83.793}{Rev. Mod. Phys. \textbf{83}, 793-836 (2011)},
[arXiv:1102.4014 [gr-qc]].
	
\bibitem{Weinberg:1972kfs}
S.~Weinberg,
Gravitation and Cosmology: Principles and Applications of the General Theory of Relativity,
John Wiley and Sons, 1972,
ISBN 978-0-471-92567-5, 978-0-471-92567-5.
	
\bibitem{Regge:1957td}
T.~Regge and J.~A.~Wheeler,
Stability of a Schwarzschild singularity,
\href{\doibase 10.1103/PhysRev.108.1063}{Phys. Rev. \textbf{108}, 1063-1069 (1957)}.
	
\bibitem{Zerilli:1970se}
F.~J.~Zerilli,
Effective potential for even parity Regge-Wheeler gravitational perturbation equations,
\href{\doibase 10.1103/PhysRevLett.24.737}{Phys. Rev. Lett. \textbf{24}, 737-738 (1970)}.
	
\bibitem{Moncrief:1974gw}
V.~Moncrief,
Odd-parity stability of a Reissner-Nordstrom black hole,
\href{\doibase 10.1103/PhysRevD.9.2707}{Phys. Rev. D \textbf{9}, 2707-2709 (1974)}.
	
\bibitem{Moncrief:1974ng}
V.~Moncrief,
Stability of Reissner-Nordstrom black holes,
\href{\doibase 10.1103/PhysRevD.10.1057}{Phys. Rev. D \textbf{10}, 1057-1059 (1974)}.
	
\bibitem{Teukolsky:1972my}
S.~A.~Teukolsky,
Rotating black holes - separable wave equations for gravitational and electromagnetic perturbations,
\href{\doibase 10.1103/PhysRevLett.29.1114}{Phys. Rev. Lett. \textbf{29}, 1114-1118 (1972)}.
	
\bibitem{Mashhoon}
B.~Mashhoon, in Proceedings of the Third Marcel Grossmann Meeting on Recent Developments of General Relativity, edited by H. Ning, p. 599, Amsterdam, 1983, North-Holland.
	
\bibitem{Schutz:1985km}
B.~F.~Schutz and C.~M.~Will,
BLACK HOLE NORMAL MODES: A SEMIANALYTIC APPROACH,
\href{\doibase 10.1086/184453}{Astrophys. J. Lett. \textbf{291}, L33-L36 (1985)}.

\bibitem{Iyer:1986np}
S.~Iyer and C.~M.~Will,
Black Hole Normal Modes: A {WKB} Approach. 1. Foundations and Application of a Higher Order {WKB} Analysis of Potential Barrier Scattering,
\href{\doibase doi:10.1103/PhysRevD.35.3621}{Phys. Rev. D \textbf{35} (1987), 3621}.

\bibitem{Konoplya:2003ii}
R.~A.~Konoplya, Quasinormal behavior of the d-dimensional Schwarzschild black hole and higher order WKB approach,
\href{\doibase doi:10.1103/PhysRevD.68.024018}{Phys. Rev. D \textbf{68}, 024018 (2003)},
[arXiv:gr-qc/0303052 [gr-qc]].

\bibitem{Matyjasek:2017psv}
J.~Matyjasek and M.~Opala,
Quasinormal modes of black holes. The improved semianalytic approach,
\href{\doibase doi:10.1103/PhysRevD.96.024011}{Phys. Rev. D \textbf{96}, no.2, 024011 (2017)},
[arXiv:1704.00361 [gr-qc]].

\bibitem{Konoplya:2019hlu}
R.~A.~Konoplya, A.~Zhidenko and A.~F.~Zinhailo,
Higher order WKB formula for quasinormal modes and grey-body factors: recipes for quick and accurate calculations,
\href{\doibase doi:10.1088/1361-6382/ab2e25}{Class. Quant. Grav. \textbf{36}, 155002 (2019)},
[arXiv:1904.10333 [gr-qc]].
	
\bibitem{Cho:2011sf}
H.~T.~Cho, A.~S.~Cornell, J.~Doukas, T.~R.~Huang, and W.~Naylor,
A New Approach to Black Hole Quasinormal Modes: A Review of the Asymptotic Iteration Method,
\href{\doibase 10.1155/2012/281705}{Adv. Math. Phys. \textbf{2012}, 281705 (2012)},
[arXiv:1111.5024 [gr-qc]].
	
\bibitem{Motl:2003cd}
L.~Motl and A.~Neitzke,
Asymptotic black hole quasinormal frequencies,
\href{\doibase 10.4310/ATMP.2003.v7.n2.a4}{Adv. Theor. Math. Phys. \textbf{7}, 307-330 (2003)},
[arXiv:hep-th/0301173].
	
\bibitem{Horowitz:1999jd}
G.~T.~Horowitz and V.~E.~Hubeny,
Quasinormal modes of AdS black holes and the approach to thermal equilibrium,
\href{\doibase 10.1103/PhysRevD.62.024027}{Phys. Rev. D \textbf{62}, 024027 (2000)},
[arXiv:hep-th/9909056].
	
\bibitem{Berti:2009wx}
E.~Berti, V.~Cardoso, and P.~Pani,
Breit-Wigner resonances and the quasinormal modes of anti-de Sitter black holes,
\href{\doibase 10.1103/PhysRevD.79.101501}{Phys. Rev. D \textbf{79}, 101501 (2009)},
[arXiv:0903.5311 [gr-qc]].
	
\bibitem{Leaver:1985ax}
E.~W.~Leaver,
An Analytic representation for the quasi normal modes of Kerr black holes,
\href{\doibase 10.1098/rspa.1985.0119}{Proc. Roy. Soc. Lond. A \textbf{402}, 285-298 (1985)}.
	
\bibitem{Rovelli:2004tv}
C.~Rovelli,
Quantum gravity,
\href{\doibase 10.1017/CBO9780511755804}{Univ. Pr., 2004}.
	
\bibitem{Modesto:2009ve}
L.~Modesto and I.~Premont-Schwarz,
Self-dual Black Holes in LQG: Theory and Phenomenology,
\href{\doibase 10.1103/PhysRevD.80.064041}{Phys. Rev. D \textbf{80}, 064041 (2009)},
[arXiv:0905.3170 [hep-th]].
	
\bibitem{Modesto:2008im}
L.~Modesto,
Semiclassical loop quantum black hole,
\href{\doibase 10.1007/s10773-010-0346-x}{Int. J. Theor. Phys. \textbf{49}, 1649-1683 (2010)}, [arXiv:0811.2196 [gr-qc]].
	
\bibitem{Alesci:2011wn}
E.~Alesci and L.~Modesto,
Particle Creation by Loop Black Holes,
\href{\doibase 10.1007/s10714-013-1656-0}{Gen. Rel. Grav. \textbf{46}, 1656 (2014)},
[arXiv:1101.5792 [gr-qc]].
	
\bibitem{Dasgupta:2012nk}
A.~Dasgupta,
Entropy Production and Semiclassical Gravity,
\href{\doibase 10.3842/SIGMA.2013.013}{SIGMA \textbf{9}, 013 (2013)},
[arXiv:1203.5119 [gr-qc]].
	
\bibitem{Barrau:2014yka}
A.~Barrau, C.~Rovelli and F.~Vidotto,
Fast Radio Bursts and White Hole Signals,
\href{\doibase 10.1103/PhysRevD.90.127503}{Phys. Rev. D \textbf{90}, 127503 (2014)}, [arXiv:1409.4031 [gr-qc]].

\bibitem{Hossenfelder:2012tc}
S.~Hossenfelder, L.~Modesto, and I.~Premont-Schwarz,
Emission spectra of self-dual black holes,
\href{https://arxiv.org/abs/1202.0412}{[arXiv:1202.0412 [gr-qc]]}.
	
\bibitem{Sahu:2015dea}
S.~Sahu, K.~Lochan, and D.~Narasimha,
Gravitational lensing by self-dual black holes in loop quantum gravity,
\href{\doibase doi:10.1103/PhysRevD.91.063001}{Phys. Rev. D \textbf{91}, 063001 (2015)},
[arXiv:1502.05619 [gr-qc]].
	
\bibitem{Zhu:2020tcf}
T.~Zhu and A.~Wang,
Observational tests of the self-dual spacetime in loop quantum gravity,
\href{\doibase doi:10.1103/PhysRevD.102.124042}{Phys. Rev. D \textbf{102}, 124042 (2020)},
[arXiv:2008.08704 [gr-qc]].
	
\bibitem{Yan:2022fkr}
J.~M.~Yan, Q.~Wu, C.~Liu, T.~Zhu, and A.~Wang,
Constraints on self-dual black hole in loop quantum gravity with S0-2 star in the galactic center,
\href{\doibase doi:10.1088/1475-7516/2022/09/008}{JCAP \textbf{09}, 008 (2022)},
[arXiv:2203.03203 [gr-qc]].
	
\bibitem{Chen:2011zzi}
J.~H.~Chen and Y.~J.~Wang,
Complex frequencies of a massless scalar field in loop quantum black hole spacetime, \href{\doibase 10.1088/1674-1056/20/3/030401}{Chin. Phys. B \textbf{20}, 030401 (2011)}.
	
\bibitem{Santos:2021wsw}
J.~S.~Santos, M.~B.~Cruz, and F.~A.~Brito,
Quasinormal modes of a massive scalar field nonminimally coupled to gravity in the spacetime of self-dual black hole,
\href{\doibase 10.1140/epjc/s10052-021-09884-1}{Eur. Phys. J. C \textbf{81}, 1082 (2021)},
[arXiv:2103.11212 [hep-th]].
	
\bibitem{Cruz:2015bcj}
M.~B.~Cruz, C.~A.~S.~Silva, and F.~A.~Brito,
Gravitational axial perturbations and quasinormal modes of loop quantum black holes,
\href{\doibase 10.1140/epjc/s10052-019-6565-2}{Eur. Phys. J. C \textbf{79}, 157 (2019)}, [arXiv:1511.08263 [gr-qc]].
	
\bibitem{Cruz:2020emz}
M.~B.~Cruz, F.~A.~Brito, and C.~A.~S.~Silva,
Polar gravitational perturbations and quasinormal modes of a loop quantum gravity black hole, \href{\doibase 10.1103/PhysRevD.102.044063}{Phys. Rev. D \textbf{102}, 044063 (2020)}, [arXiv:2005.02208 [gr-qc]].
	
\bibitem{Liu:2020ola}
C.~Liu, T.~Zhu, Q.~Wu, K.~Jusufi, M.~Jamil, M.~Azreg-A\"\i{}nou, and A.~Wang,
Shadow and quasinormal modes of a rotating loop quantum black hole,
\href{\doibase 10.1103/PhysRevD.101.084001}{Phys. Rev. D \textbf{101}, 084001 (2020) [erratum: Phys. Rev. D \textbf{103} (2021) no.8, 089902]}, [arXiv:2003.00477 [gr-qc]].
	
\bibitem{Momennia:2022tug}
M.~Momennia,
Quasinormal modes of self-dual black holes in loop quantum gravity,
\href{\doibase 10.1103/PhysRevD.106.024052}{Phys. Rev. D \textbf{106}, 024052 (2022)}, [arXiv:2204.03259 [gr-qc]].
	
\bibitem{Chen:2019iuo}
C.~Y.~Chen and P.~Chen,
Gravitational perturbations of nonsingular black holes in conformal gravity,
\href{\doibase 10.1103/PhysRevD.99.104003}{Phys. Rev. D \textbf{99}, 104003 (2019)}, [arXiv:1902.01678 [gr-qc]].
	
\bibitem{Bouhmadi-Lopez:2020oia}
M.~Bouhmadi-L\'opez, S.~Brahma, C.~Y.~Chen, P.~Chen, and D.~h.~Yeom,
A consistent model of non-singular Schwarzschild black hole in loop quantum gravity and its quasinormal modes,
\href{\doibase 10.1088/1475-7516/2020/07/066}{JCAP \textbf{07}, 066 (2020)},
[arXiv:2004.13061 [gr-qc]].
	
\bibitem{Cardoso:2008bp}
V.~Cardoso, A.~S.~Miranda, E.~Berti, H.~Witek, and V.~T.~Zanchin,
Geodesic stability, Lyapunov exponents and quasinormal modes,
\href{\doibase 10.1103/PhysRevD.79.064016}{Phys. Rev. D \textbf{79}, 064016 (2009)}, [arXiv:0812.1806 [hep-th]].
	
\bibitem{Gundlach:1993tp}
C.~Gundlach, R.~H.~Price, and J.~Pullin,
Late time behavior of stellar collapse and explosions: 1. Linearized perturbations, \href{\doibase doi:10.1103/PhysRevD.49.883}{Phys. Rev. D \textbf{49}, 883-889 (1994)}, [arXiv:gr-qc/9307009].

\bibitem{Konoplya:2003ii}
R.~A.~Konoplya,
Quasinormal behavior of the d-dimensional Schwarzschild black hole and higher order WKB approach,
\href{\doibase doi:10.1103/PhysRevD.68.024018}{Phys. Rev. D \textbf{68} (2003), 024018},
[arXiv:gr-qc/0303052 [gr-qc]].
		
\bibitem{Cardoso:2003vt}
V.~Cardoso, J.~P.~S.~Lemos, and S.~Yoshida,
Quasinormal modes of Schwarzschild black holes in four-dimensions and higher dimensions, \href{\doibase doi:10.1103/PhysRevD.69.044004}{Phys. Rev. D \textbf{69}, 044004 (2004)}, [arXiv:gr-qc/0309112].
	
\bibitem{Konoplya:2017wot}
R.~A.~Konoplya and Z.~Stuchl\'\i{}k, Are eikonal quasinormal modes linked to the unstable circular null geodesics?,
\href{\doibase 10.1016/j.physletb.2017.06.015}{Phys. Lett. B \textbf{771}, 597-602 (2017)}, [arXiv:1705.05928 [gr-qc]].

\bibitem{Zou:2021lkj}
Y.~Zou, M.~Wang and J.~Jing,
Test of a model coupling of electromagnetic and gravitational fields by using high-frequency gravitational waves,
\href{\doibase doi:10.1007/s11433-020-1674-0}{Sci. China Phys. Mech. Astron. \textbf{64} (2021) no.5, 250411}.

\bibitem{Guo:2022rms}
W.~D.~Guo, Q.~Tan and Y.~X.~Liu,
Gravito-Electromagnetic coupled perturbations and quasinormal modes of a charged black hole with scalar hair,
\href{\doibase https://arxiv.org/abs/2212.08784}{[arXiv:2212.08784 [gr-qc]]}.
	
\bibitem{Cardoso:2019mqo}
V.~Cardoso, M.~Kimura, A.~Maselli, E.~Berti, C.~F.~B.~Macedo, and R.~McManus,
Parametrized black hole quasinormal ringdown: Decoupled equations for nonrotating black holes, \href{\doibase 10.1103/PhysRevD.99.104077}{Phys. Rev. D \textbf{99}, 104077 (2019)}, [arXiv:1901.01265 [gr-qc]].
	
\bibitem{McManus:2019ulj}
R.~McManus, E.~Berti, C.~F.~B.~Macedo, M.~Kimura, A.~Maselli, and V.~Cardoso,
Parametrized black hole quasinormal ringdown. II. Coupled equations and quadratic corrections for nonrotating black holes,
\href{doi:10.1103/PhysRevD.100.044061}{Phys. Rev. D \textbf{100}, 044061 (2019)}, [arXiv:1906.05155 [gr-qc]].
	
\bibitem{Volkel:2022khh}
S.~H.~V\"olkel, N.~Franchini, E.~Barausse, and E.~Berti,
Constraining modifications of black hole perturbation potentials near the light ring with quasinormal modes,
\href{\doibase 10.1103/PhysRevD.106.124036}{Phys. Rev. D \textbf{106}, 124036 (2022)}, [arXiv:2209.10564 [gr-qc]].

\bibitem{Feng:2020tyc}
W.~B.~Feng, S.~J.~Yang, Q.~Tan, J.~Yang, and Y.~X.~Liu, Overcharging a Reissner-Nordstr\"om Taub-NUT regular black hole,
\href{\doibase doi:10.1007/s11433-020-1659-0}{Sci. China Phys. Mech. Astron. \textbf{64} (2021) no.6, 260411},
[arXiv:2009.12846 [gr-qc]].

\bibitem{Yifu}
Y.~F. Cai,
Remarkable effects of multiple photon spheres on black hole observations,
\href{https://doi.org/10.1007/s11433-022-2019-7}{Sci. China Phys. Mech. Astron. \textbf{65} (2022) no.12, 120431}.

\bibitem{Zeng:2021mok}
X.~X.~Zeng, K.~J.~He, and G.~P.~Li,
Effects of dark matter on shadows and rings of Brane-World black holes illuminated by various accretions,
\href{\doibase doi:10.1007/s11433-022-1896-0}{Sci. China Phys. Mech. Astron. \textbf{65} (2022) no.9, 290411},
[arXiv:2111.05090 [gr-qc]].

\bibitem{Chen:2022qrw}
Y.~Chen, G.~Guo, P.~Wang, H.~Wu, and H.~Yang, Appearance of an infalling star in black holes with multiple photon spheres,
\href{\doibase doi:10.1007/s11433-022-1986-x}{Sci. China Phys. Mech. Astron. \textbf{65} (2022) no.12, 120412},
[arXiv:2206.13705 [gr-qc]].

\bibitem{Liu:2022ruc}
X.~Liu, S.~Chen, and J.~Jing,
Polarization distribution in the image of a synchrotron emitting ring around a regular black hole,
\href{\doibase 10.1007/s11433-022-1946-2}{Sci. China Phys. Mech. Astron. \textbf{65} (2022) no.12, 120411},
[arXiv:2205.00391 [gr-qc]].

\bibitem{Wei:2021bwy}
S.~W.~Wei, Y.~Q.~Wang, Y.~X.~Liu, and R.~B.~Mann,
Observing dynamic oscillatory behavior of triple points among black hole thermodynamic phase transitions,
\href{\doibase doi:10.1007/s11433-021-1706-2}{Sci. China Phys. Mech. Astron. \textbf{64} (2021) no.7, 270411},
[arXiv:2102.00799 [gr-qc]].

\bibitem{Cai:2021sag}
R.~G.~Cai,
Oscillatory behaviors near a black hole triple point,
\href{\doibase doi:10.1007/s11433-021-1738-5}{Sci. China Phys. Mech. Astron. \textbf{64} (2021) no.9, 290432}.

\bibitem{Song:2020arr}
S.~Song, H.~Li, Y.~Ma, and C.~Zhang,
Entropy of black holes with arbitrary shapes in loop quantum gravity,
\href{\doibase doi:10.1007/s11433-021-1770-3}{Sci. China Phys. Mech. Astron. \textbf{64} (2021) no.12, 120411},
[arXiv:2002.08869 [gr-qc]].
\end{thebibliography}
\end{document}